\documentclass{article}

\usepackage{PRIMEarxiv}

\usepackage[utf8]{inputenc} 
\usepackage[T1]{fontenc}    
\usepackage{hyperref}       
\usepackage{url}            
\usepackage{booktabs}       
\usepackage{amsfonts}       
\usepackage{nicefrac}       
\usepackage{microtype}      
\usepackage{lipsum}
\usepackage{fancyhdr}       

\usepackage{times}
\usepackage[dvipdfmx]{graphicx}
\usepackage{bm}
\usepackage{comment}
\usepackage{amsmath}
\renewcommand{\figurename}{Fig.}

\title{Emergence of sensory attenuation based upon the free-energy principle
}

\author 
{\centerline{Hayato Idei$^{1\ast}$, Wataru Ohata$^{2}$, Yuichi Yamashita$^{1}$, Tetsuya Ogata$^{3}$ and Jun Tani$^{2\ast}$}\\
\\
\centerline{\normalsize{$^{1}$Department of Information Medicine, National Center of Neurology and Psychiatry, Tokyo, Japan}}\\
\centerline{\normalsize{$^{2}$Cognitive Neurorobotics Research Unit, Okinawa Institute of Science and Technology, Okinawa, Japan}}\\
\centerline{\normalsize{$^{3}$Department of Intermedia Art and Science, Waseda University, Tokyo, Japan}}\\
\\
\centerline{\normalsize{$^\ast$Correspondence to: jun.tani@oist.jp or idei@ncnp.go.jp}}
}

\begin{document}
\maketitle
\centerline{This manuscript has been accepted for publication in \it{Scientific Reports}.}

\begin{abstract}
The brain attenuates its responses to self-produced exteroceptions (e.g., we cannot tickle ourselves). Is this phenomenon, known as sensory attenuation, enabled innately, or acquired through learning? Here, our simulation study using a multimodal hierarchical recurrent neural network model, based on variational free-energy minimization, shows that a mechanism for sensory attenuation can develop through learning of two distinct types of sensorimotor experience, involving self-produced or externally produced exteroceptions. For each sensorimotor context, a particular free-energy state emerged through interaction between top-down prediction with precision and bottom-up sensory prediction error from each sensory area. The executive area in the network served as an information hub. Consequently, shifts between the two sensorimotor contexts triggered transitions from one free-energy state to another in the network via executive control, which caused shifts between attenuating and amplifying prediction-error-induced responses in the sensory areas. This study situates emergence of sensory attenuation (or self-other distinction) in development of distinct free-energy states in the dynamic hierarchical neural system.
\end{abstract}


\section*{Introduction}
The brain couples its structure with the outside world via sensorimotor experiences~\cite{Clark2013}. %
A posteriori development of neural processing gradually forms our perceptual phenomena, with a priori nature determined by genes. %
It yields well-defined self-experience and helps to confidently situate self in relation to others. %
In general, we face two primary types of sensorimotor experience, self-movements and sensory events in the outside world, which may or may not be correlated. %
Recognition of the difference is thought to underlie self-other distinctions, sense of self, and subsequent attribution of agency~\cite{Braun2018,Legaspi2019}, but the difference cannot be known a priori. %
The brain may acquire the capacity to modulate neural responses via sensorimotor learning, depending on the context. %
\par
A phenomenon called sensory attenuation, is recognized as one of the bases of the sense of self, especially the sense of agency~\cite{Dewey2014,Braun2018}. %
Sensory attenuation refers to an experience in which an exteroception produced by a self-movement is less salient than one produced externally~\cite{Weiskrantz1971}. %
A perfect example is the difficulty of tickling ourselves. %
In addition, the ability to ignore visual changes during eye or head movements is thought to help maintain stability of the visual scene. %
At the neural level, sensory attenuation is observed as reduced brain responses, such as blood oxygen level-dependent (BOLD) responses, especially in sensory areas~\cite{Blakemore1998,Pamela2008,Arikan2019}.  %
A line of research shows that sensory attenuation is associated with predicted sensorimotor correlation and is diminished by temporal or spatial mismatch between a movement and the resultant exteroception (e.g., a temporal delay or a trajectory rotation)~\cite{Blakemore2000,Bays2005,Kilteni2018}. %
There are some suggestions of underlying neural functions, such as an efference copy of the motor command~\cite{Wolpert1995,Blakemore2000} and neuromodulation (e.g., dopaminergic transmission)~\cite{Brown2013}. %
However, despite intensive work for decades, a fundamental question remains unclear: Is sensory attenuation enabled innately or is it acquired through learning? %
\par
Here, we provide a computational explanation, suggesting that a mechanism for sensory attenuation can be self-organized through learning. %
We focus on the following factors. %
If a self-movement and a sensory event in the outside world are not correlated, the resultant proprioception and exteroception occur separately. %
The brain may efficiently use individual sensory areas to represent these individual sensations. %
On the other hand, if proprioception and exteroception are correlated, they can be best explained as being generated by the same latent cause, i.e., self, rather than being generated by two independent latent states, i.e., self and other. %
In that case, if the brain can develop a predictive model about the spatio-temporal sensorimotor correlation, it is reasonable to represent the sensorimotor coupling using a multimodal association area, and neural responses in individual sensory areas can be attenuated. %
That is, through sensorimotor learning, the brain may develop the capacity to predictively modulate sensory-information flow and processing inside hierarchical neural networks, depending on the context, i.e., self generated versus other generated. %
To test this hypothesis, we conducted a robotic simulation experiment using a variational recurrent neural network (RNN) model based on Bayesian brain theory or free-energy minimization~\cite{Clark2013,Friston2010,Adams2013,Ahmadi2019,WOhata2020}. %
\section*{Results}
\subsection*{Computational model}
Our daily sensorimotor behavior generally has some sort of regularity, e.g., a set-point of body posture and dynamic movement patterns, and randomness, e.g., free movement with fluctuations~\cite{Inoue2020}. %
Therefore, we considered how a robot agent controlled by an RNN develops sensory attenuation through such spontaneous sensorimotor experiences (Fig. 1). %
A robot repeatedly generated a random behavior (imprecise movement) and then returned to the set-posture (precise movement) within five seconds (20 time steps for the robot), where the robot's motion and a movement of an external red object were correlated (self-produced context) or uncorrelated (externally produced context). %
The received sensations were 3-dimensional proprioception for joint angles and 2-dimensional exteroception for the position of the object. %
Previous brain imaging studies show that sensory attenuation involves hierarchical interactions among multiple brain regions, including sensory areas, an association area (parietal area), and a higher cognitive area (prefrontal or supplementary motor area)~\cite{Haggard2004,Wolpe2016,Boehme2019}. %
The RNN inside the robot has a corresponding hierarchical structure, referred to as sensory (exteroceptive and proprioceptive), association, and executive areas (Fig. 1a). %
The hierarchical RNN is a generative model that predicts sensations as well as inferring hidden causes of sensations via variational free-energy minimization~\cite{Ahmadi2019,WOhata2020}. %
Each area has individual latent variables $\bm{z_{t}}$ representing each level of belief about the hidden cause of sensation in the form of Gaussian distributions. %
Each latent variable has prior $\bm{z^{p}_{t}}$ and posterior $\bm{z^{q}_{t}}$ distributions that correspond to estimated hidden causes before and after observing sensations, respectively. %
The priors and posteriors of latent variables are time-varying in sensory and association areas, while the executive area maintains a constant posterior state during sequential prediction generation within a time window (Supplementary Fig. S1-2). %
We assumed that the executive area controls (and switches) sequential patterns of lower-level network behavior, like the prefrontal or supplementary motor area in a biological brain~\cite{Eagleman2004,Leek2009}. %
At each time step, the RNN generates predictions $\bm{\hat{x}_{t}}$ about the next exteroceptive and proprioceptive sensations $\bm{x_{t}}$ from latent variables via recurrent units $\bm{d_{t}}$, representing dynamics of the environment. %
Variational free-energy $F_{t}$ can be calculated as the sum of prediction error and Kullback-Leibler divergence between the posterior and prior at each $l$th network level. %
\begin{eqnarray}
    \displaystyle F_{t} =\underbrace{ \frac{1}{2} \left(\bm{x_{t}}-\bm{\hat{x}_{t}} \right)^{2}}_{\rm{Prediction \: error}} + \sum_{l=1}^{3}W^{(l)}\underbrace{D_{KL}[q(\bm{z_{t}^{(l)}}|\bm{e_{t:T}})||p(\bm{z_{t}^{(l)}}|\bm{d_{t-1}^{(l)}})]}_{\rm{Influence \: of \: prior \: beliefs}}.
\end{eqnarray}
This describes integration of sensory information, i.e., prediction error $\bm{e_{t}}$, and prior knowledge, i.e., prior belief $\bm{z^{p}_{t}}$, upon a posterior $\bm{z^{q}_{t}}$ update, although synaptic weights are also updated during the learning process. %
The decomposition of variational free-energy in the equation (1) can also be thought of in terms of accuracy (prediction error) and complexity (the divergence between posterior and prior beliefs). %
A posterior is only adjusted to minimize prediction errors under the influence of the prior. %
A weak prior (with low precision) leads to a sensitive posterior update in response to a prediction error, while a strong prior (with high precision) leads to discounting of the prediction error. %
In other words, prediction errors have a greater effect on posterior beliefs in the presence of a weak, i.e., imprecise prior.
The strength of the prior in each area is parameterized by the standard deviation of a Gaussian prior that is dynamically modulated to minimize the path integral of free-energy. %
In other words, we equip the model with the capacity to infer differences in prior precision in a way that lends the influence of prior beliefs a context sensitivity. %
From the available physiological evidence, it may engage local GABAergic mechanisms or neuromodulators, e.g., acetylcholine or dopamine~\cite{Yu2005,Corlett2010,Friston2010}. %
In addition, we introduce a hyper-parameter called the meta-prior $W^{(l)}$, which controls the meta-level balance between the prediction error and the divergence between the posterior and the prior at each network level. %
Previous studies have shown that a large meta-prior causes an intrinsically strong prior, whereas a small meta-prior causes an intrinsically weak prior~\cite{Ahmadi2019,WOhata2020}. %
The inclusion of meta-priors may seem superfluous, given that the precision, i.e., inverse variance, of various beliefs are inferred adaptively. %
However, meta-priors offer an opportunity to model a failure to infer context sensitive changes in precision. %
This particular failure of sensory attenuation may be an important explanation for several neuropsychiatric conditions: see below. %
We assumed that the meta-prior represents an innately equipped characteristic in the biological brain. %
In the baseline model, we set the meta-prior to the same value for all network levels ($W^{(1)}=W^{(2)}=W^{(3)}=0.005$) to avoid bias in the prediction-error flow. %

\subsection*{Emergence of sensory attenuation}
This experiment comprises learning and test phases. %
In the learning phase, the robot learned to reproduce the two types of sensorimotor patterns (Fig. 1b). %
Target sensorimotor sequences were prepared in advance from human spontaneous behaviors through manual manipulations of a physical robot. %
We prepared 24 training sets for each set of self-produced and externally produced contexts. %
We ensured that each of the 24 movement patterns of the robot was the same for both contexts (see Fig. 1b and ``Learning" section in the Methods). %
The RNN updated posteriors and synaptic weights to minimize the path integral of free-energy (Supplementary Fig. S1). %
\par
In the test phase, the trained robot was required to generate actions by itself and to recognize an environmental context by updating only posteriors to minimize the free-energy, with fixed synaptic weights (Fig. 2a and Supplementary Fig. S2). %
Action generation of the robot was performed by proportional-integral-derivative (PID) control. %
A PID controller receives proprioceptive predictions as target joint angles and it changes the joint angles (proprioception) to minimize the error between the current state and target. %
In this regard, the PID controller implements active inference under the free-energy principle by realizing posterior predictions about movement trajectories~\cite{Adams2013}. %
Fig. 2b-e shows an example of a test trial (Supplementary Video S1). %
The test trial consists of a self-produced context during time steps 0-100 and an externally produced context during time steps 100-200. %
\par
In the self-produced context, the robot moved the external object by itself, and the object position matched the robot's hand position (Fig. 2e). %
The trained RNN successfully reproduced the stochastic property of learned sensorimotor sequences by modulating the association-level latent mean $\bm{\mu^{(2)}}$, such that the estimated sigma $\bm{\sigma^{(2)}}$ was high during random behavior generation, but low when returning to the set-posture (Fig. 2c). %
Here, $\bm{\mu}$ and $\bm{\sigma}$ correspond to the mean and standard deviation of a Gaussian posterior or prior. %
In other words, they play the role of sufficient statistics of approximate posterior beliefs and prior beliefs that are optimized during learning. %
Modulation of the latent mean in exteroceptive and proprioceptive areas $\bm{\mu^{(1)}}$ was small, with low estimated sigma $\bm{\sigma^{(1)}}$ (Fig. 2d), suggesting reduced prediction-error flow into sensory areas with high precision in its prior. %
These results indicate that the RNN minimized prediction errors produced by self-movements mainly by adjusting the posterior in the association area, since the posterior in sensory areas cannot be easily adjusted because of the strong prior. %
\par
Then, at time step 100, the environment was shifted to externally produced contexts, where the object position was given from test data that were not used in the learning phase. %
The object position and hand position became uncorrelated (Fig. 2e). %
The environmental change caused a stepwise change in the executive-level latent state $\bm{\mu^{(3)}}$ (Fig. 2b). %
At the same time, in sensory areas, modulation in the latent mean $\bm{\mu^{(1)}}$ was amplified and sigma $\bm{\sigma^{(1)}}$ increased, showing periodic modulation (Fig. 2d), like the association-level latent state. %
This shows that the RNN minimized prediction errors from externally produced sensations by adjusting the posterior in sensory areas, as well as in the association area. %
\par
Collectively, the hierarchical RNN attenuated neural responses in sensory areas for the self-produced context and amplified them for the externally produced context by proactively controlling precision of the prior at each network level in which the posterior in the executive area worked as the information hub for switching the lower-level precision structure and the prediction error flow (Fig. 3). %
Furthermore, free-energy converged into different states for distinct sensorimotor contexts (Fig. 3). %
This suggests that a particular free-energy minimum was developed for each sensorimotor context and the transition of the free-energy state in the network, induced by the abrupt context shift, underlay the qualitative change in the free-energy minimization. %
\par
For quantitative analysis, we prepared 10 trained networks with different initial synaptic weights and conducted 8 test trials for each trained network. %
Fig. 4a shows the change in the sensory-level posterior mean $\bm{\mu^{(1),q}}$ per time step for the two contexts. %
In this study, it is referred to as a sensory-level posterior response. %
A paired t-test reported that the posterior response was significantly smaller in the self-produced context than in the externally produced context ($t(9)=-3.38, p=0.0082$). %
In addition, a paired t-test showed that the sensory-level prior sigma $\bm{\sigma^{(1),p}}$ in the self-produced context was significantly lower than that in the externally produced context ($t(9)=-3.32, p=0.0089$) (Fig. 4b). %
\par
Furthermore, we analyzed how attenuation of neural responses in sensory areas developed during the learning process. %
The RNN first increased sensory-level posterior responses to reconstruct target sensorimotor sequences (Fig. 4c). %
Then, sensory-level posterior responses in the self-produced context were gradually attenuated in both exteroceptive and proprioceptive areas. 
The sensory-level prior sigma diminished more in the self-produced context than in the externally produced context through the learning process (Fig. 4d). %
We confirmed that posterior responses in the association area were similar in self-produced and externally produced contexts (Supplementary Fig. S3a-b), indicating reduced total neural response in the self-produced context. %
This sensory attenuation was accompanied by recognition of both contexts in the executive area (Supplementary Fig. S3c-d). %
These results demonstrate emergence of sensory attenuation through a learning process via free-energy minimization. %
Additional analyses indicated that development of sensory attenuation was diminished by removing neurons in the association or executive area (Supplementary Fig. S4), suggesting the importance of a higher-level representation of sensorimotor correlation. %
\par
In addition, we investigated effects by modulation of the meta-prior at each network level (Supplementary Fig. S5-7). %
In particular, a small meta-prior in sensory areas or a large meta-prior in the association area led to a deficit in attenuation of the sensory-level posterior response, as well as the sensory-level prior sigma, in the self-produced context. %
This suggests that innately decreased prediction-error flow into the association area compared to sensory areas disrupted development of sensory attenuation. %

\section*{Discussion}
The current model study, using a hierarchically organized variational RNN, illustrated the possibility that a sensory attenuation mechanism can develop through learning. %
In the learning task, the robot alternately repeated imprecise movement (random behavior) and precise movement (returning to the set-posture) in both self-produced and externally produced contexts. %
The RNN developed a hierarchical generative model about how proprioceptive and exteroceptive inputs are generated from the latent causes and also represented the stochastic property by dynamically modulating precisions of hierarchical latent variables, which are individually allocated to each area of the RNN. %
We found that for dealing with two distinct types of sensorimotor contexts, namely self-produced and externally produced contexts, the network developed two distinct free-energy states (minima) through learning, wherein each free-energy state corresponds to each sensorimotor context (Fig. 3 and 4c-d). %
In the developed network, the top-down and bottom-up pathways functioned as follows. %
In the top-down pathway, the posterior in the executive area predicted the prior precision of the association area and sensory areas. %
In the bottom-up pathway, prediction error determined the posterior of sensory areas, the association area, and the executive area under the constraint of the prior precision expected at each area by the top-down pathway. %
Here, we see that the top-down pathway and the bottom-up pathway created a closed circuit in which the posterior in the executive area served as the information hub. %
In the closed circuit, a particular free-energy state corresponded to the characteristic top-down precision control and bottom-up prediction-error flow inside the hierarchical RNN.%
\par
In the self-produced context, sensory attenuation was achieved by minimizing the free-energy to the corresponding free-energy state. %
In this context, the posterior in the executive area developed to a particular value, and induced high prior precision in sensory areas and low prior precision in the association area, which resulted in less adaptation in its posterior in sensory areas and more adaptation in the association area using the bottom-up error signal. %
Less adaptation in the posterior in sensory areas corresponds to sensory attenuation. %
The detail mechanism is as below. %
The hierarchical RNN recognized that proprioceptive and exteroceptive inputs were generated from the same latent cause, i.e., self, and the latent cause, including its stochastic property, was represented in the higher-level areas (association area and executive area). %
In this case, sensory areas were not needed to represent the latent cause at all. %
Indeed, Fig. 2 shows that the RNN can reconstruct sensory inputs based on the dynamic modulation of latent variables only in the association area. %
Then, high precision (nearly 0 sigma) of latent variables in the sensory areas is thought to be developed to minimize variational free-energy by reducing prediction error due to random sampling and reducing KL divergence between the posterior and the prior. %
\par
In an externally produced context, the process was the other way around, wherein the posterior in sensory areas adapted greatly because of the low precision in its prior regulated by the executive area. %
This corresponds to sensory amplification. %
In this context, proprioceptive and exteroceptive inputs were generated from individual causes, i.e., self, and other. %
The hierarchical RNN needed to represent not only the higher-level context, i.e., externally produced context, using the association and executive areas, but also unresolved lower-level information due to individual causes using individual sensory areas. %
That is why, precisions of latent variables in the sensory areas as well as the association area were modulated dynamically. %
\par
Furthermore, the error induced by the change of the sensorimotor context flowed bottom-up to the executive area and determined the posterior in the executive area with its prior set to a neutral value. %
It triggered the transition from one free-energy state to another, i.e., from sensory attenuation to sensory amplification and verse-versa. %
In short, precision structures for sensory attenuation in self-produced contexts and sensory amplification in externally produced contexts were self-organized in one hierarchical RNN and were switched via executive control. %
This suggests that the hierarchical RNN developed a type of functionality of switching between quasi-deterministic and highly stochastic dynamics in each local area, in which sensory attenuation was characterized by quasi-deterministic processing (nearly 0 sigma) in the sensory areas and highly stochastic processing in the association area, while sensory amplification was characterized by highly stochastic processing in both the sensory areas and the association area. %
This sort of development and transitions of distinct free-energy states provides insights into how perceptual phenomena emerge from dynamic brain-body-environment interaction in the face of uncertainty. %
\par
Sensory attenuation observed in our model is an emergent property based on variational free-energy minimization, rather than a consequence guaranteed by the equations used in the proposed model. %
Indeed, a small meta-prior in sensory areas led to a deficit in development of sensory attenuation, in which prediction error and free-energy for training data were smaller than those of the baseline model (see Supplementary Fig. S5 and S8). %
This shows that sensory attenuation was not necessarily developed even if proprioceptive and exteroceptive sensory inputs are correlated and the model precisely reconstructs the sensory inputs based on variational free-energy minimization. %
Instead, development of sensory attenuation depends on a generative model of the world self-organized by the hierarchical RNN. %
In particular, the following factors suggest that development of sensory attenuation required a sort of abstract-level recognition that proprioceptive and exteroceptive inputs are generated from the same latent cause, i.e., sensorimotor coupling, which was achieved by a balanced interaction between the top-down prediction with precision and the bottom-up prediction error. %
First, reduced numbers of latent variables in the association area or executive area disrupted development of sensory attenuation (Supplementary Figure S4). %
Second, a large meta-prior in the association area as well as a small meta-prior in the sensory areas, i.e., reduced prediction-error flow into the association area, led to a deficit in development of sensory attenuation (Supplementary Fig. S5-6). %
These show the importance of higher-level structure learning for developing sensory attenuation. %
Furthermore, the difference in the dimensions of proprioception (3-dimensional joint angle) and exteroception (2-dimensional object position) required an abstract-level representation for recognizing the sensorimotor coupling. %
Such an abstraction of sensorimotor information may be a reasonable consequence given the efficient coding, but it is not necessarily obvious given the self-organization by the hierarchical RNN itself through learning. %
This kind of emergent phenomenon self-organized through learning has not thoroughly been investigated in previous studies based on the free-energy principle. %
Thus, we think that this complex property emphasizes the significance of our neurorobotic approach. %
\par
There have been prior model proposals to account for sensory attenuation. %
One proposal~\cite{Brown2013} postulates that sensory attenuation is caused by reducing the precision of the prediction error bottom-up to the sensory area during movement by following the free-energy principle. %
This model, however, does not explain the involvement of the higher executive area, as evidenced by \cite{Haggard2004,Wolpe2016}. %
We confirmed that removal of the executive area diminished the development of sensory attenuation (Supplementary Fig. S4b), emphasizing the contribution of the frontal function. %
This is consistent with biological studies suggesting that signals from the frontal area, such as the supplementary motor area, predictably control the relative precision or intensity of
sensations. %
Their disruption causes diminished sensory attenuation~\cite{Haggard2004,Pynn2013}. %
Furthermore, the previous model intermixes two phenomena: sensory attenuation and sensory gating. %
Sensory attenuation compares the intensity of self-produced sensations with externally produced sensations, or the distinction between the self and others. %
On the other hand, sensory gating refers to a suppression process in which exteroceptions feel weaker during movement than at rest~\cite{Kilteni2022}. %
In our learning experiment, movements of the robot in self-produced contexts and externally produced contexts were the same, avoiding the confusion with sensory gating. %
In this sense, our model considered only sensory attenuation (self-other distinction). %
\par
According to another leading hypothesis, the pathway of an efference copy of the motor command is thought to originate from top-down signals~\cite{Haggard2004,Pynn2013}. %
In contrast, our model suggests that signals from the executive (frontal) area may represent predictive signals for controlling prediction-error flow inside the hierarchical network, rather than a copy of the motor command. %
We showed that the functionality, in which top-down signals hierarchically control bottom-up prediction-error flow inside the network (that modulates top-down signals), can be self-organized through learning. %
\par
The perspective that the sensory attenuation mechanism is a consequence of learning instead of an innate function, may be indirectly supported by a recent study suggesting that a target stimulus of sensory attenuation can be adaptively changed through rapid learning~\cite{Kilteni2019}. %
In addition, our result (Fig. 4c) explains increased sensory attenuation with age in adults~\cite{Wolpe2016}. %
Furthermore, our model suggests that proprioception, as well as exteroception, can be attenuated when a self-movement produces exteroception. %
In fact, a neuroimaging study observes less cerebellar activity when a movement produces a tactile stimulus, than when it does not~\cite{Blakemore1998}. %
\par
There have been some recent advances in our understanding of movement-related suppression processes, including the finding of a difference between sensory attenuation and sensory gating. %
One of the discussions is about a difference between ``physiological sensory attenuation" and ``perceptual sensory attenuation". %
A recent study suggests that ``physiological sensory attenuation" and ``perceptual sensory attenuation" have different neurophysiological correlates~\cite{Palmer2016}. %
In a recent study, ``physiological sensory attenuation" was measured as a decrease in the amplitude of primary and secondary components of the somatosensory evoked potential (SEP) by comparing it during movement and at rest; thus, it may be related with sensory gating. %
On the other hand, ``perceptual sensory attenuation" was measured in a force-matching paradigm and was suggested to be related with a decrease in prediction-error-related neural activity, such as gamma-oscillatory activity. %
Importantly, ``perceptual sensory attenuation" negatively correlated with scores of delusional ideation (a measure of schizotypy) while no significant correlations were found between attenuation of SEP components and scores of delusional ideation. %
In our experiment, we focus on an attenuation of prediction-error-related response of the posterior; thus, our model may explain ``perceptual sensory attenuation" and its neurocomputational mechanism. %
In addition, our model of sensory attenuation suggests that a decrease in prediction-error-related activity and ``perceptual sensory attenuation" may represent self-other distinction, which has implications for mechanisms underlying delusional ideation. %
\par
In addition, a recent study demonstrated an enhancement, not suppression, of the intensity of predicted action outcomes by investigating effects of action on intensity of tactile stimulus reported by participants in the force-judgment paradigm~\cite{Thomas2022}. %
First, the authors replicated typical findings that self-produced tactile stimuli are rated as less intense than externally produced stimuli when an active contact with a button generates a tactile stimulus. %
However, this effect reversed when there was no active finger contact with a button. %
In additional experiments, they controlled the predictability of tactile action outcomes and found that expected events were perceived more, not less, intensely than unexpected events. %
Note that this additional experiment compared cases with more predictable and less predictable outcomes produced by own action. %
Since this comparison did not compare stimuli produced by own action and other's action, this additional experimental result does not conflict with our claim. %
These results in the previous study~\cite{Thomas2022} may show that when sensory attenuation does not appear, an action can enhance expected touch. %
This enhancement effect is consistent with the basic Bayesian idea that a predictive brain focuses on precise (predictable) stimuli to ignore uninformative (uncertain) noise. %
In this sense, the previous study suggests that a theory of sensory attenuation requires an additional thought beyond the basic Bayesian or active inference framework. %
Here, our model provides a particular mechanism for sensory attenuation beyond the basic Bayesian theory. %
Specifically, our results suggest that sensory attenuation requires an abstract-level recognition that proprioceptive and exteroceptive inputs are generated from the same latent cause, i.e., self, which is developed through context-sensitive optimization of hierarchical latent variables.
This emphasizes the importance of developmental aspects of hierarchical predictive processing. %
\par
Alterations in development of sensory attenuation were induced by modulation of the meta-prior, a hyper-parameter determining the intrinsic strength of the prior compared to the prediction error at each network level. %
In the basic Bayesian model, the practitioner must manually set the prior, including its precision, to compute the posterior. %
However, in our model, the prior in each network area is epigenetically self-organized through learning, with the influence of the higher meta-level parameter, i.e., the meta-prior. %
We assumed that the meta-prior is an innate characteristic determining developmental features of the prior and prediction-error flow in the biological brain, although its neural substrate remains unspecified. %
In our experiment, a large meta-prior in the association area, i.e., an intrinsically strong prior in the association area, led to reduced sensory attenuation, i.e., a weak prior in sensory areas, while a small meta-prior in the association area, i.e., an intrinsically weak prior in the association area, did not impair development of sensory attenuation (Supplementary Fig. S6). %
These results may provide insights into relationships between strong prior hypotheses and reduced sensory attenuation in schizophrenia~\cite{Blakemore2000,Brown2013,Powers2017,Corlett2019}  and between weak prior hypotheses and normal sensory attenuation in autism spectrum disorder~\cite{Blakemore2006,Lawson2014,Haker2016,Palmer2017,Finnemann2021}. %
It will be important to seek corresponding neurological evidence of the proposed computational mechanism for development of sensory attenuation and to explore its relevance to psychiatric disorders in future studies. %
\par
In summary, we have shown that sensory attenuation is an emergent property of free-energy minimization in active inference. %
Sensory attenuation is a ubiquitous phenomenon in psychology and psychophysics. %
Furthermore, it is receiving increasing attention in computational psychiatry, in which a failure of sensory attenuation may explain several neuropsychiatric conditions, ranging from Parkinson's disease to schizophrenia and autism~\cite{Brown2013,Lawson2014}. %
Technically, we have simulated behavior in terms of active inference, which can be thought of as a generalization of control and planning as inference. %
Crucially, we trained our robots to perform active inference by optimizing connection weights ($\bm{w}$) and adaptive variables ($\bm{a}$) in hierarchically organized RNNs, which played the role of a hierarchical generative model. %
The implicit learning of connection strengths can be thought of as ``amortization" or ``learning to infer". %
In other words, by minimizing (the path or time integral of) variational free-energy, robots were able to learn the mapping from sensory inputs to posterior beliefs about the causes of their sensory inputs. %
These posterior beliefs recognize the context in which they are operating and generate proprioceptive predictions that are realized by a PID controller. %
Because certain beliefs concern precision of inferred sensorimotor trajectories, this enables a context sensitive optimization of certain precisions that underwrite sensory attenuation. %
Put simply, when  robots recognized that the proprioceptive and exteroceptive inputs are best explained by externally generated movement, they reduced the precision afforded representations of self generated sensations. %
Conversely, when proprioceptive and exteroceptive sensory inputs were best explained by self-generated sensations, precision of these explanations increased, thereby reducing the influence of sensory prediction errors. %
This is the basis of sensory attenuation. %
Note that this context-sensitive optimization of precision, i.e., encoding of uncertainty, is an emergent property of minimizing the path integral of free-energy. %
In other words, sensory attenuation emerges from a generative model that provides the best explanation for proprioceptive and exteroceptive inputs, when accumulating free-energy or model evidence (a.k.a., marginal likelihood) over time. %

\section*{Methods}
\subsection*{Neural network model}
To simulate development of sensory attenuation, we utilized a predictive-coding-inspired variational recurrent neural network (PV-RNN), which represents a generative process of sensation from a hidden cause in the environment, based on the free-energy principle (FEP)~\cite{Ahmadi2019,WOhata2020}. %
It consists of sensory (exteroceptive and proprioceptive), association, and executive areas in which there are deterministic neurons and latent (stochastic) neurons. %
Latent neurons represent belief about the cause of sensation as Gaussian distributions (for simplicity). %
Each latent state has prior and posterior probability distributions that correspond to estimated hidden causes before and after observing sensations, respectively. %
Based on the latent states, the PV-RNN generates predictions about next sensations in a top-down way. %
Here, deterministic neurons transform latent states into sensory predictions via synaptic connections that represent relationships between sensations and their causes. %
The PV-RNN uses a multiple timescale RNN (MTRNN)~\cite{Yamashita2008} as the transformation function. %
An RNN represents temporal processing of the brain in that neural activity is determined by the past history of neural states. %
Owing to their capacity to learn to reproduce complex dynamic behaviors, RNNs have been used in computational modeling and developmental neurorobotic studies to understand cortical processing and cognitive functions, including psychiatric symptoms~\cite{Yamashita2008,Yamashita2012,Idei2018,Idei2020,Idei2021,Finkelstein2021}. %
In addition, MTRNNs have a multiple timescale property in neural activation, which enables them to represent a temporal hierarchy in the environment, as observed in the biological brain~\cite{Newell2001}. %
By using a PV-RNN, we can learn complex sensorimotor behaviors that have both temporal regularity and stochasticity~\cite{Ahmadi2019,WOhata2020}. %
To perform spontaneous behaviors like daily human behaviors, a neural network-controlling robot was required to process short-term random movements and a long-term periodic pattern, as in the experiment in which the multiple timescale property is thought to facilitate sensorimotor learning. %
In the following sections, we describe the mathematical details of top-down prediction generation and bottom-up parameter updates by the PV-RNN (Supplementary Fig. S1-2). %

\subsubsection*{Prediction generation}
Prediction generation is performed in a top-down way through the network hierarchy. %
The internal state $h_{t,i}^{(s)}$ and output $d_{t,i}^{(s)}$ of the $i$th deterministic neuron of the $s$th target sequence at time step $t$ $\left(t \ge 1\right)$ is calculated as 
\begin{eqnarray}
 h_{t,i}^{(s)} = \left \{
 \begin{array}{ll}
    \displaystyle \frac{1}{\tau}\Biggl(\sum_{j\in I_{\rm Ad}}w_{ij}d_{t-1,j}^{(s)}
    +\sum_{j\in I_{\rm Az}}w_{ij}z_{t,j}^{(s)}+\sum_{j\in I_{\rm Cz}}w_{ij}z_{t,j}^{(s)}+b_{i}\Biggr)+\Biggl(1-\frac{1}{\tau}\Biggr)h_{t-1,i}^{(s)}
    \hspace{16truept}\left(i \in I_{\rm Ad}\right),
    \vspace{10truept}\\ \\ 
    \displaystyle \frac{1}{\tau}\Biggl(\sum_{j\in I_{\rm Ed}}w_{ij}d_{t-1,j}^{(s)}
    +\sum_{j\in I_{\rm Ez}}w_{ij}z_{t,j}^{(s)}+\sum_{j\in I_{\rm Ad}}w_{ij}d_{t,j}^{(s)}+b_{i}\Biggr)+\Biggl(1-\frac{1}{\tau}\Biggr)h_{t-1,i}^{(s)}
    \hspace{16truept}\left(i \in I_{\rm Ed}\right),
    \vspace{10truept}\\ \\
     \displaystyle \frac{1}{\tau}\Biggl(\sum_{j\in I_{\rm Pd}}w_{ij}d_{t-1,j}^{(s)}
    +\sum_{j\in I_{\rm Pz}}w_{ij}z_{t,j}^{(s)}+\sum_{j\in I_{\rm Ad}}w_{ij}d_{t,j}^{(s)}+b_{i}\Biggr)+\Biggl(1-\frac{1}{\tau}\Biggr)h_{t-1,i}^{(s)}
    \hspace{16truept}\left(i \in I_{\rm Pd}\right). \end{array} \right. \hspace{-40truept}\nonumber\\
\end{eqnarray}
\begin{eqnarray}
  &d_{t,i}^{(s)} = \tanh{\left(h_{t,i}^{(s)}\right)}
  &\hspace{10truept}\left(i \in I_{\rm Ed},I_{\rm Pd},I_{\rm Ad}\right).
\end{eqnarray}
Here, $I_{\rm Ed}$, $I_{\rm Pd}$, and $I_{\rm Ad}$ are index sets of deterministic neurons in the exteroceptive area, proprioceptive area, and association area, respectively. %
$I_{\rm Ez}$, $I_{\rm Pz}$, $I_{\rm Az}$, and $I_{\rm Cz}$ are index sets of latent neurons in the exteroceptive, proprioceptive, association, and executive areas, respectively. %
$w_{ij}$ is the weight of the synaptic connection from the $j$th neuron to the $i$th neuron; $z_{t,j}^{(s)}$ is the output of $j$th latent (posterior) neuron at time step $t$; $\tau$ is the time constant of the neuron; and $b_{i}$ is the bias of the $i$th neuron. %
A deterministic neuron with a small time constant $\tau$ has a tendency to change its activity rapidly, while that with a large time constant has a tendency to change its activity slowly. %
We set the initial internal states of the deterministic neurons $h_{0,i}^{(s)} \left(i \in I_{\rm Ed},I_{\rm Pd},I_{\rm Ad}\right)$ to $0$ ($d_{0,i}^{(s)}$ is also $0$). %
\par
The latent variable $\bm{z}$ in each area is assumed to follow a multivariate Gaussian distribution with a diagonal covariance matrix, meaning $z_{t,i}^{(s)}$ and $z_{t,j}^{(s)}$ are independent ($i,j \in I_{\rm Ez},I_{\rm Pz},I_{\rm Az} \land i \neq j$). %
Here, the mean $\mu_{t,i}^{(s),p}$ and sigma (standard deviation) $\sigma_{t,i}^{(s),p}$ of the prior distribution $p(z_{t,i}^{(s)})$ in the exteroceptive, proprioceptive, and association areas are calculated from the previous deterministic state (prior experience) of the same area. %
\begin{eqnarray}
 \displaystyle p(z_{t,i}^{(s)}) = p(z_{t,i}^{(s)}|d_{t-1,j}^{(s)})=\mathcal{N}(z_{t,i}^{(s)};\mu_{t,i}^{(s),p},\sigma_{t,i}^{(s),p}).
\end{eqnarray}
\begin{eqnarray}
  &\mu_{t,i}^{(s),p} = \tanh{\left( \sum_{j}w_{ij}d_{t-1,j}^{(s)} \right)}, \vspace{10truept}\\
  &\sigma_{t,i}^{(s),p} = \exp{\left( \sum_{j}w_{ij}d_{t-1,j}^{(s)} \right)}.
\end{eqnarray}
Here, $\left(i \in I_{\rm Ez} \land j \in I_{\rm Ed}\right) \lor \left(i \in I_{\rm Pz} \land j \in I_{\rm Pd}\right) \lor \left(i \in I_{\rm Az} \land j \in I_{\rm Ad}\right)$. %
Note by optimizing the weights $\bm{w}$, with respect to the path integral of free-energy, we are effectively optimizing prior beliefs about sensorimotor contingencies and contexts. %
This can be regarded as a form of structure learning through experience. %
The executive area has a prior distribution as $\mathcal{N}(0,1)$ only at the initial step ($t=1$) because it has a constant posterior state during sequential prediction generation, with the objective of assigning a specific executive-level posterior to each target sequence. %
\par
The posterior distribution in each area is calculated as,
\begin{eqnarray}
 q(z_{t,i}^{(s)}|\bm{e_{t:T^{(s)}}^{(s)}}) = \left \{
 \begin{array}{ll}
    \displaystyle q(z_{t-1,i}^{(s)}|\bm{e_{t-1:T^{(s)}}^{(s)}}) = q(z_{1,i}^{(s)}|\bm{e_{1:T^{(s)}}^{(s)}})=\mathcal{N}(z_{1,i}^{(s)};\mu_{1,i}^{(s),q},\sigma_{1,i}^{(s),q})
    \hspace{16truept}\left(i \in I_{\rm Cz}\right),
    \vspace{10truept}\\ \\ 
    \displaystyle \mathcal{N}(z_{t,i}^{(s)};\mu_{t,i}^{(s),q},\sigma_{t,i}^{(s),q})
    \hspace{16truept}\left(i \in I_{\rm Ez}, I_{\rm Pz},I_{\rm Az}\right). \end{array} \right. \hspace{-40truept}\nonumber\\
\end{eqnarray}
\begin{eqnarray}
  &\mu_{t,i}^{(s),q} = \tanh{\left(a_{t,i}^{(s),\mu} \right)}, \vspace{10truept}\\
  &\sigma_{t,i}^{(s),q} = \exp{\left(a_{t,i}^{(s),\sigma} \right)}, \vspace{10truept}\\
  &z_{t,i}^{(s)} = \mu_{t,i}^{(s),q} + \sigma_{t,i}^{(s),q} \times \epsilon.
\end{eqnarray}
Here, $T^{(s)}$ represents the length of the $s$th target sequence. %
$\bm{a}$ is the adaptive internal state of neurons representing posterior distributions and it is updated at each time step and for each target sequence during the learning process (or each time step through online inference). %
Note that equations (8) and (9) mean that the adaptive variables implicitly encode both posterior expectations and precision, such that optimizing the adaptive variables with respect to variational free-energy implicitly optimizes posterior expectations and the posterior confidence or precision afforded those expectations. %
Adaptive variables $\bm{a_{t}}$ are determined by prediction error signals $\bm{e_{t:T}}$ propagated by a back-propagation-through-time (BPTT) algorithm, meaning that the posterior of the latent state can be considered a prediction-error-related neural state. %
Adaptive variables $\bm{a}$ are initialized by the corresponding initial internal states of the neurons representing prior distributions before the learning or inference process. %
Based on the posterior calculation, the latent state $z_{t,i}^{(s)}$ is obtained by sampling $\epsilon$ from $\mathcal{N}(0,1)$. %
\par
Finally, predictions about exteroceptive and proprioceptive sensations are individually generated from exteroceptive and proprioceptive areas, respectively.
\begin{eqnarray}
 \hat{x}_{t,i}^{(s)} = \left \{
 \begin{array}{ll}
    \displaystyle \tanh{\left(\sum_{j\in I_{\rm Ed}}w_{ij}d_{t,j}^{(s)} \right)}
    \hspace{34truept}\left(i \in I_{\rm Eo} \right),
     \vspace{10truept}\\ \\
     \displaystyle \tanh{\left(\sum_{j\in I_{\rm Pd}}w_{ij}d_{t,j}^{(s)} \right)}
     \hspace{34truept} \left(i \in I_{\rm Po}\right). \end{array} \right. \hspace{-40truept}\nonumber\\
\end{eqnarray}
Here, $I_{\rm Eo}$ and $I_{\rm Po}$ are index sets of output neurons for exteroceptive and proprioceptive predictions, respectively. %

\subsubsection*{Parameter updates via free-energy minimization}
The concept of FEP~\cite{Friston2010} derives from the fundamental fact that self-organizing biological agents must maintain a limited repertoire of sensory states to remain alive, e.g., a human stays on the ground, not in the sea. %
Based on information theory, this notion can be formulated as suppression of the surprise (or the negative log-evidence) for sensations $\bm{x}$ over time. %
In the PV-RNN, the surprise over all time steps and target sequences can be written as:
\begin{eqnarray}
    \displaystyle &&\sum_{s \in I_{\rm S}}-\log p(\bm{x_{1:T^{(s)}}^{(s)}}) = \sum_{s \in I_{\rm S}}\left[-\log \int p(\bm{x_{1:T^{(s)}}^{(s)}},\bm{z_{1:T^{(s)}}^{(s)}})d\bm{z_{1:T^{(s)}}^{(s)}}\right] \\
    &=& \sum_{s \in I_{\rm S}}\left[-\log \int \prod_{t=1}^{T^{(s)}} \left[ p(\bm{x_{t}^{(s)}}|\bm{d_{t}^{(s)}})p(\bm{z_{t}^{(s)}}|\bm{d_{t-1}^{(s)}})\right] d\bm{z_{1:T^{(s)}}^{(s)}}\right].
\end{eqnarray}
Here, $I_{\rm S}$ denotes the index set of target sequences. %
However, surprise cannot be directly evaluated by the agent because it needs to know all hidden states $\bm{z}$ of the environment that cause sensations, as described on the right side of the equation (12). %
Here, FEP introduces a tractable quantity, free-energy, that bounds the surprise, and minimization of the surprise is replaced by minimization of the free-energy. %
The bound of the surprise in the PV-RNN can be derived by utilizing Jensen's inequality for a concave function $f$: $f(E[x])\geq E[f(x)]$. %
For clarity, summation over target sequences is temporarily omitted in the following equations. %
Then, equation (13) can be deformed as follows by introducing a dummy distribution for $\bm{z_{1}}$, $q(\bm{z_{1}})$. %
\begin{eqnarray}
    \displaystyle &&-\log p(\bm{x_{1:T}}) \notag\\&& = -\log \int_{\bm{z_{1}}}d\bm{z_{1}}q(\bm{z_{1}}) \int_{\bm{z_{2:T}}}d\bm{z_{2:T}} \frac{1}{q(\bm{z_{1}}) }\prod_{t=1}^{T} \left[p(\bm{x_{t}}|\bm{d_{t}})p(\bm{z_{t}}|\bm{d_{t-1}})\right] \\
    &&\leq - \int_{\bm{z_{1}}}d\bm{z_{1}}q(\bm{z_{1}}) \log \int_{\bm{z_{2:T}}}d\bm{z_{2:T}} \frac{1}{q(\bm{z_{1}}) }\prod_{t=1}^{T} \left[p(\bm{x_{t}}|\bm{d_{t}})p(\bm{z_{t}}|\bm{d_{t-1}})\right].
\end{eqnarray}
In equation (15), we use Jensen's inequality for a logarithmic function. %
The same procedure is done for $t=2:T$. %
\begin{eqnarray}
    \displaystyle -\log p(\bm{x_{1:T}}) \leq - \int_{\bm{z_{1}}}d\bm{z_{1}}q(\bm{z_{1}}) \cdots \int_{\bm{z_{T}}}d\bm{z_{T}}q(\bm{z_{T}}) \log \prod_{t=1}^{T} \left[\frac{p(\bm{x_{t}}|\bm{d_{t}})p(\bm{z_{t}}|\bm{d_{t-1}})}{q(\bm{z_{t}})} \right] \\
    = - \int_{\bm{z_{1}}}d\bm{z_{1}}q(\bm{z_{1}}) \cdots \int_{\bm{z_{T}}}d\bm{z_{T}}q(\bm{z_{T}}) \sum_{t=1}^{T} \left[ \underbrace{\log p(\bm{x_{t}}|\bm{d_{t}})}_{\rm{A}} - \underbrace{\log \frac{q(\bm{z_{t}})}{p(\bm{z_{t}}|\bm{d_{t-1}})}}_{\rm{B}} \right].
\end{eqnarray}
The first term in expression (17) is the expected negative log-likelihood under $q$ given that $\bm{d_{t}}$ depends on $\bm{z_{1:t}}$. %
\begin{eqnarray}
    \displaystyle \rm{A} &=& - \int_{\bm{z_{1}}}d\bm{z_{1}}q(\bm{z_{1}}) \cdots \int_{\bm{z_{T}}}d\bm{z_{T}}q(\bm{z_{T}}) \sum_{t=1}^{T} \log p(\bm{x_{t}}|\bm{d_{t}}) \\
    &=& - \sum_{t=1}^{T} E_{q(\bm{z_{1:t}})} \left[\log p(\bm{x_{t}}|\bm{d_{t}})\right].
\end{eqnarray}
In addition, the second term can be deformed into forms of Kullback-Leibler divergence (KLD) between $q(\bm{z_{t}})$ and $p(\bm{z_{t}}|\bm{d_{t-1}})$ while being careful to ensure that $\bm{d_{0}}$ is independent from $\bm{z_{1:T}}$. %
\begin{eqnarray}
    \displaystyle &\rm{B}& = \int_{\bm{z_{1}}}d\bm{z_{1}}q(\bm{z_{1}}) \cdots \int_{\bm{z_{T}}}d\bm{z_{T}}q(\bm{z_{T}}) \sum_{t=1}^{T} \log \frac{q(\bm{z_{t}})}{p(\bm{z_{t}}|\bm{d_{t-1}})} \\
    &=& \int_{\bm{z_{1}}}d\bm{z_{1}}q(\bm{z_{1}})\log \frac{q(\bm{z_{1}})}{p(\bm{z_{1}}|\bm{d_{0}})} \notag\\ &&+ \int_{\bm{z_{1}}}d\bm{z_{1}}q(\bm{z_{1}}) \cdots \int_{\bm{z_{T}}}d\bm{z_{T}}q(\bm{z_{T}}) \sum_{t=2}^{T} \log \frac{q(\bm{z_{t}})}{p(\bm{z_{t}}|\bm{d_{t-1}})} \\
    &=& D_{KL}[q(\bm{z_{1}})||p(\bm{z_{1}}|\bm{d_{0}})] +  \sum_{t=2}^{T} E_{q(\bm{z_{1:t-1}})} \left[ D_{KL}[q(\bm{z_{t}})||p(\bm{z_{t}}|\bm{d_{t-1}})]  \right].
\end{eqnarray}
A dummy distribution $q(\bm{z_{t}})$ can be replaced by the posterior distribution determined by the back-propagated prediction error $q(\bm{z_{t}}|\bm{e_{t:T}})$. %
Thus, using equations (19) and (22), the bound of the surprise can be written as,
\begin{equation}
\begin{split}
    -\log p(\bm{x_{1:T}}) \leq &- \sum_{t=1}^{T} E_{q(\bm{z_{1:t}}| \bm{e_{t:T}})} \left[\log p(\bm{x_{t}}|\bm{d_{t}})\right] + D_{KL}[q(\bm{z_{1}} | \bm{e_{1:T}} )||p(\bm{z_{1}}|\bm{d_{0}})] + \\
    &\sum_{t=2}^{T} E_{q(\bm{z_{1:t-1}} | \bm{e_{t-1:T}})} \left[ D_{KL}[q(\bm{z_{t}} | \bm{e_{t:T}})||p(\bm{z_{t}}|\bm{d_{t-1}})]  \right].
\end{split}
\end{equation}
In the experiment, we resolve calculation of the expectations of negative log-likelihood and KLD under $q$ by considering single sampling to reduce computational costs. %
\begin{eqnarray}
    \displaystyle -\log p(\bm{x_{1:T}}) &\leq& \sum_{t=1}^{T} \left( -\log p(\bm{x_{t}}|\bm{d_{t}}) + D_{KL}[q(\bm{z_{t}} | \bm{e_{t:T}})||p(\bm{z_{t}}|\bm{d_{t-1}})] \right).
\end{eqnarray}    
Eventually, by introducing the meta-prior and considering the summation over different target sequences, the free-energy $F$ (the bound of the surprise) in the PV-RNN is formulated as:
\begin{eqnarray}
    \displaystyle F &=& \sum_{s \in I_{\rm S}} \sum_{t=1}^{T^{(s)}} F_{t}^{(s)} \\
    F_{t}^{(s)} &=& - \underbrace{ \log p(\bm{x_{t}^{(s)}}|\bm{d_{t}^{(s)}})}_{\rm{Accuracy}} + \notag\\ &&\sum_{l=1}^{3}W^{(l)} \underbrace{ \left( D_{KL}[q(\bm{z_{t}^{(l),(s)}} | \bm{e_{t:T}^{(s)}})||p(\bm{z_{t}^{(l),(s)}}|\bm{d_{t-1}^{(l),(s)}})]  \right)}_{\rm{Complexity}}.
\end{eqnarray}
Here, $W^{(l)}$ denotes the meta-prior at $l$th network level. %
The first term in expression (26) (negative accuracy term) is the negative log-likelihood. %
For simplicity, we assume that each sensory state follows a Gaussian distribution with unit variance. %
Then, the first term becomes just the prediction error between the real $\bm{x_{t}^{(s)}}$ and predicted $\bm{\hat{x}_{t}^{(s)}}$ sensations (plus a constant term, omitted here). %
\begin{eqnarray}
    \displaystyle -{\rm{Accuracy}} = \sum_{i \in I_{\rm Eo} \lor I_{\rm Po}} \frac{1}{2} \left(x_{t,i}^{(s)}-\hat{x}_{t,i}^{(s)} \right)^{2}.
\end{eqnarray}
This assumption sets the precision of the prediction error to a constant value and makes it easy to consider the relative precision of prior beliefs compared to prediction errors. %
In the experiment, we divided the accuracy term by the dimension of each exteroceptive and proprioceptive sensation. %
\par
On the other hand, the second term (complexity term) is the KLD between the posterior and prior distributions of the latent variables. %
Note that variables of the posterior are updated through both the accuracy and complexity terms, but those of the prior are updated only through the complexity term. %
Therefore, the complexity term represents only the influence of prior beliefs, which are controlled by the meta-prior $W$. %
Under the assumption that the prior and posterior distributions follow a multivariate Gaussian distribution with a diagonal covariance matrix, as described above, the KLD is computed analytically as~\cite{Ahmadi2019,WOhata2020},
\begin{eqnarray}
    \displaystyle {\rm{Complexity}} = \sum_{i} \left(\log\frac{\sigma_{t,i}^{(s),p}}{\sigma_{t,i}^{(s),q}} + \frac{(\mu_{t,i}^{(s),p} - \mu_{t,i}^{(s),q})^2 + (\sigma_{t,i}^{(s),q})^2}{2(\sigma_{t,i}^{(s),p})^2} - \frac{1}{2} \right).
\end{eqnarray}
Here, $i \in I_{\rm Ez} \lor I_{\rm Pz}$ (if $l=1$), $i \in I_{\rm Az}$ (if $l=2$), and $i \in I_{\rm Cz}$ (if $l=3$). %
In the experiment, we divided the complexity term by the dimension of latent variables for each area. %
Note that if $l=3$ (executive area), the complexity term exits only at $t=1$, although we write the equation in a general way for simplicity. %
\par
In the learning phase (Supplementary Fig. S1), synaptic weights $\bm{w}$ and adaptive variables $\bm{a}$ are updated to minimize the free-energy over all time steps and target sequences as,
\begin{equation}
  F  =  \sum_{s \in I_{\rm S}} \sum_{t=1}^{T^{(s)}}F_{t}^{(s)}.
\end{equation}
In the test phase after learning (Supplementary Fig. S2), only adaptive variables $\bm{a}$ are updated, while synaptic weights are fixed. %
In this phase, the free-energy within a short time window $H$ is summed as
\begin{equation}
  F  = \sum_{t^{'}=t-H+1}^{t}F_{t^{'}}.
\end{equation}
Using the summed free-energy, adaptive variables $\bm{a_{t-H+1:t}}$ within the time window of all areas are updated, whereas the time window slides as the network time step $t$ is incremented. %
 \par
In both learning and test phases, we used the Adam optimizer~\cite{Kingma2017} for parameter updates, where the partial derivative of the free-energy with respect to each parameter is calculated by the BPTT algorithm. %

\subsection*{Experimental environment}
We set a 3-axis robotic arm in a simulated square space $[-1,1]\times[-1,1]$. %
The lengths of the robot's links are 0.1, 0.3, and 0.5. %
Each joint angle was limited to range from $0$ to $\pi$ [rad] and normalized to range from $-0.8$ to $0.8$ to match the range of the neural network output. %
In addition, the PV-RNN receives the 2-dimensional position of a red object as an exteroceptive sensation. %
During task execution, the robot receives 5-dimensional sensations every 250 ms. %

\subsection*{Learning}
The PV-RNN learned to reproduce target sensorimotor sequences prepared in advance. %
First, we recorded 24 sequences of joint angles while the experimenter manually manipulated the left arm of a physical robot (Rakuda, developed by Robotis). %
For each sequence, the experimenter performed 10 repetitions generating a random movement and returning to the set posture within five seconds (20 time steps). %
Therefore, the length of each sequence is 200 time steps. %
We used joint angles from left shoulder pitch, left shoulder roll, and the left elbow of Rakuda as 3-dimensional proprioception data of the simulated robot arm. %
Next, we prepared exteroception data for a self-produced context by setting the object position as the hand position of the simulated arm robot that was calculated by forward kinematics using the recorded joint angle data. %
In this fashion, we obtained 24 target sequences for self-produced contexts in which exteroceptive and proprioceptive sensations are correlated. %
Finally, we prepared target sequences for externally produced contexts by shuffling the combination of exteroceptive and proprioceptive sequences. %
By doing this, we obtained 24 target sequences for externally produced contexts in which exteroceptive and proprioceptive sensations are not correlated. %
This shuffling procedure ensures that total numbers of changes in sensations are the same for self-produced and externally produced contexts in the learning phase. %
The PV-RNN learned to reproduce the prepared 48 training datasets via free-energy minimization. %

\subsection*{Online inference}
We prepared an additional 8 exteroceptive sequences as test data that were used in externally produced contexts in the test phase. %
Before a test trial, initial states of adaptive variables $\bm{a_{1}}$ in all areas were set to the median values obtained for 24 training datasets of the self-produced context developed through learning. %
Based on the initial posterior states corresponding to self-produced contexts, the robot first moved the object by itself during time steps 0-100. %
The robot controlled its joint angles via active inference using the PID controller, for which proprioceptive predictions were used as target joint angles. %
Then, during time steps 100-200, the environmental context was shifted to the externally produced context, although the robot kept generating spontaneous behaviors by itself via active inference. %
In the externally produced context, the object position was set from test data. %
The goal of the robot was to flexibly recognize the environmental change by updating adaptive variables via free-energy minimization. %
The online inference process was performed based on an interaction between top-down prediction generation and bottom-up posterior updates. %
In the top-down prediction generation process, the PV-RNN generated sensory predictions $\bm{\hat{x}_{t-H+1:t}}$ corresponding to time steps from $t-H+1$ to $t$, based on the posterior of latent states $\bm{z_{t-H+1:t}}$. %
In the bottom-up modulation process, the free-energy at each time step in time window $H$ was calculated using prediction errors for exteroception and proprioception, for which target sensations were the real object position and joint angles from $t-H+1$ to $t$. %
Adaptive variables $\bm{a_{t-H+1:t}}$ in the time window were updated to minimize the free-energy summed over time steps, and sensory predictions within the time window were re-generated using the updated posterior states. %
By repeating top-down prediction generation and bottom-up posterior updates for a certain duration, the PV-RNN generated the prior of latent states for time step $t+1$. %
The generated prior was used to initialize the posterior for time step $t+1$, and predictions about sensations for time step $t+1$ were generated from the posterior. %
Using proprioceptive predictions for time step $t+1$ as the target joint angles, the robot moved the joint angles using the PID controller. %
At the same time, the robot received exteroceptive sensations at time step $t+1$. %
After that, the robot's time step was incremented and the online inference process was performed for the newly received sensations. %
This inference process, in which recognition and prediction in the past are reconstructed from current sensations, corresponds to a ``postdiction" process. %

\subsection*{Parameter settings}
The dimension of latent variables $\bm{z}$ in the exteroceptive and proprioceptive areas was 1, and that in the association area was 3. %
Therefore, the total dimension of latent neurons in sensory and association areas was the same as the sensory dimension. %
The dimension of latent variables in executive area was 1. %
A preliminary experiment showed that a smaller number of latent neurons in the association area led to a lower level of sensory attenuation, supporting the idea that sensory attenuation is a consequence of representing sensorimotor correlation in the association area (Supplementary Fig. S4a). %
In addition, we confirmed that no executive-level latent state resulted in a highly decreased level of sensory attenuation, suggesting an important role of the executive-level latent state (Supplementary Fig. S4b). %
The numbers of deterministic neurons in the exteroceptive, proprioceptive, and association areas were all 15. %
In a preliminary experiment, we evaluated development of sensory attenuation for settings of 10, 15, or 20 deterministic neurons and confirmed that the setting of 15 neurons showed the largest sensory attenuation (Supplementary Fig. S9a). %
We set the time constant $\tau$ to half the number of deterministic neurons (8 neurons) to 2 and that of the other neurons (7 neurons) to 4, as the simplest multiple timescale setting. %
We set the same multiple timescale property for both the sensory and association areas. %
This is because we thought that the PV-RNN controlled which levels of the network hierarchy should be used to represent sensations, depending on the context. %
Actually, the experimental results show that the PV-RNN mainly used the association area in the self-produced context and both the sensory and association areas in an externally produced context, in which the timescale property included in sensations was the same for the two contexts. %
In a preliminary experiment, we confirmed that the multiple timescale setting led to greater sensory attenuation compared to the single timescale setting $\tau=2$ for all deterministic neurons (Supplementary Fig. S9b). %
Synaptic weights were initialized with random values using the default method implemented by Pytorch. 
Biases of deterministic neurons were initialized with and fixed to random values following a Gaussian distribution $N(0,10)$, for which the variance of biases is close to the firing threshold variability found in biological neurons~\cite{Azouz2000}, as well as the optimal value in a spiking neural network model~\cite{Mejias2012} and a recurrent neural network model~\cite{Idei2020}. %
We trained 10 networks with different initial synaptic weights for each hyper-parameter setting for quantitative evaluations. %
In the learning phase, parameters including synaptic weights $\bm{w}$ and adaptive variables $\bm{a}$ were updated $200,000$ times with the Adam optimizer. %
We used the same parameter setting of Adam as in the original paper: $\alpha= 0.001$ (learning rate), $\beta_{1}=0.9$, and $\beta_{2}=0.999$. %
In the test phase, adaptive variables $\bm{a}$ were updated 50 times at each time step for a time window of length $H = 10$ with $\alpha= 0.09$. %
We chose the optimal parameter setting in the test phase from the combinations of $\alpha=\{0.001, 0.005, 0.01, 0.03, 0.05, 0.07, 0.09, 0.1, 0.2, 0.3, 0.4, 0.5\} \times H=\{10, 15\}$ by evaluating levels of prediction error in the baseline meta-prior setting ($W^{(1)}=W^{(2)}=W^{(3)}=0.005$). %

\subsection*{Statistical analysis}
We used paired t-tests for statistical analyses of network behaviors, such as the posterior response and the level of the prior sigma. %
All statistical tests were two-tailed, and the significance level was set at $p<0.05$. %
The current study conducted an original unprecedented computational simulation experiment. %
Thus, it is difficult to estimate the effect size, and no statistical methods were used to pre-determine sample size. %
Considering the high reproducibility of computational simulation, we set the minimum sample size that seemed statistically testable (10 samples). %
Indeed, even for 10 samples, paired t-tests reported clear differences between self-produced context and externally produced context (e.g., $t(9)=-3.38; p=0.0082$ for posterior response and $t(9)=-3.32; p=0.0089$ for prior sigma). %
Therefore, we concluded that a larger sample size would not have significantly influenced our main result. %
Data analyses were conducted using R software (version 3.3.2). %


\begin{thebibliography}{10}
\urlstyle{rm}
\expandafter\ifx\csname url\endcsname\relax
  \def\url#1{\texttt{#1}}\fi
\expandafter\ifx\csname urlprefix\endcsname\relax\def\urlprefix{URL }\fi
\expandafter\ifx\csname doiprefix\endcsname\relax\def\doiprefix{DOI: }\fi
\providecommand{\bibinfo}[2]{#2}
\providecommand{\eprint}[2][]{\url{#2}}
\providecommand\JournalTitle[1]{#1}

\bibitem{Clark2013}
\bibinfo{author}{Clark, A.}
\newblock \bibinfo{journal}{\bibinfo{title}{Whatever next? predictive brains,
  situated agents, and the future of cognitive science}}.
\newblock {\emph{\JournalTitle{Behavioral and Brain Sciences}}}
  \textbf{\bibinfo{volume}{36}}, \bibinfo{pages}{181--204},
  \doiprefix\url{10.1017/S0140525X12000477} (\bibinfo{year}{2013}).

\bibitem{Braun2018}
\bibinfo{author}{Braun, N.} \emph{et~al.}
\newblock \bibinfo{journal}{\bibinfo{title}{The senses of agency and ownership:
  A review}}.
\newblock {\emph{\JournalTitle{Frontiers in Psychology}}}
  \textbf{\bibinfo{volume}{9}}, \bibinfo{pages}{535},
  \doiprefix\url{10.3389/fpsyg.2018.00535} (\bibinfo{year}{2018}).

\bibitem{Legaspi2019}
\bibinfo{author}{Legaspi, R.} \& \bibinfo{author}{Toyoizumi, T.}
\newblock \bibinfo{journal}{\bibinfo{title}{A bayesian psychophysics model of
  sense of agency}}.
\newblock {\emph{\JournalTitle{Nature Communications}}}
  \textbf{\bibinfo{volume}{10}}, \bibinfo{pages}{4250},
  \doiprefix\url{10.1038/s41467-019-12170-0} (\bibinfo{year}{2019}).

\bibitem{Dewey2014}
\bibinfo{author}{Dewey, J.~A.} \& \bibinfo{author}{Knoblich, G.}
\newblock \bibinfo{journal}{\bibinfo{title}{Do implicit and explicit measures
  of the sense of agency measure the same thing?}}
\newblock {\emph{\JournalTitle{PLOS ONE}}} \textbf{\bibinfo{volume}{9}},
  \bibinfo{pages}{e110118}, \doiprefix\url{10.1371/journal.pone.0110118}
  (\bibinfo{year}{2014}).

\bibitem{Weiskrantz1971}
\bibinfo{author}{Weiskrantz, L.}, \bibinfo{author}{Elliot, J.} \&
  \bibinfo{author}{Darlington, C.}
\newblock \bibinfo{journal}{\bibinfo{title}{Preliminary observations on
  tickling oneself}}.
\newblock {\emph{\JournalTitle{Nature}}} \textbf{\bibinfo{volume}{230}},
  \bibinfo{pages}{598--599}, \doiprefix\url{10.1038/230598a0}
  (\bibinfo{year}{1971}).

\bibitem{Blakemore1998}
\bibinfo{author}{Blakemore, S.-J.}, \bibinfo{author}{Wolpert, D.~M.} \&
  \bibinfo{author}{Frith, C.~D.}
\newblock \bibinfo{journal}{\bibinfo{title}{Central cancellation of
  self-produced tickle sensation}}.
\newblock {\emph{\JournalTitle{Nature Neuroscience}}}
  \textbf{\bibinfo{volume}{1}}, \bibinfo{pages}{635--640},
  \doiprefix\url{10.1038/2870} (\bibinfo{year}{1998}).

\bibitem{Pamela2008}
\bibinfo{author}{B^^c3^^a4^^c3^^9f, P.}, \bibinfo{author}{Jacobsen, T.} \&
  \bibinfo{author}{Schr^^c3^^b6ger, E.}
\newblock \bibinfo{journal}{\bibinfo{title}{Suppression of the auditory n1
  event-related potential component with unpredictable self-initiated tones:
  Evidence for internal forward models with dynamic stimulation}}.
\newblock {\emph{\JournalTitle{International Journal of Psychophysiology}}}
  \textbf{\bibinfo{volume}{70}}, \bibinfo{pages}{137--143},
  \doiprefix\url{10.1016/j.ijpsycho.2008.06.005} (\bibinfo{year}{2008}).

\bibitem{Arikan2019}
\bibinfo{author}{Arikan, B.~E.} \emph{et~al.}
\newblock \bibinfo{journal}{\bibinfo{title}{Perceiving your hand moving: Bold
  suppression in sensory cortices and the role of the cerebellum in the
  detection of feedback delays}}.
\newblock {\emph{\JournalTitle{Journal of Vision}}}
  \textbf{\bibinfo{volume}{19(14)}}, \doiprefix\url{10.1167/19.14.4}
  (\bibinfo{year}{2019}).

\bibitem{Blakemore2000}
\bibinfo{author}{Blakemore, S.-J.}, \bibinfo{author}{Wolpert, D.} \&
  \bibinfo{author}{Frith, C.}
\newblock \bibinfo{journal}{\bibinfo{title}{Why can’t you tickle yourself?}}
\newblock {\emph{\JournalTitle{NeuroReport}}} \textbf{\bibinfo{volume}{11}},
  \bibinfo{pages}{R11--R16} (\bibinfo{year}{2000}).

\bibitem{Bays2005}
\bibinfo{author}{Bays, P.~M.}, \bibinfo{author}{Wolpert, D.~M.} \&
  \bibinfo{author}{Flanagan, J.~R.}
\newblock \bibinfo{journal}{\bibinfo{title}{Perception of the consequences of
  self-action is temporally tuned and event driven}}.
\newblock {\emph{\JournalTitle{Current Biology}}}
  \textbf{\bibinfo{volume}{15}}, \bibinfo{pages}{1125--1128},
  \doiprefix\url{10.1016/j.cub.2005.05.023} (\bibinfo{year}{2005}).

\bibitem{Kilteni2018}
\bibinfo{author}{Kilteni, K.}, \bibinfo{author}{Andersson, B.~J.},
  \bibinfo{author}{Houborg, C.} \& \bibinfo{author}{Ehrsson, H.~H.}
\newblock \bibinfo{journal}{\bibinfo{title}{Motor imagery involves predicting
  the sensory consequences of the imagined movement}}.
\newblock {\emph{\JournalTitle{Nature Communications}}}
  \textbf{\bibinfo{volume}{9}}, \bibinfo{pages}{1--9},
  \doiprefix\url{10.1038/s41467-018-03989-0} (\bibinfo{year}{2018}).

\bibitem{Wolpert1995}
\bibinfo{author}{Wolpert, D.~M.}, \bibinfo{author}{Ghahramani, Z.} \&
  \bibinfo{author}{Jordan, M.~I.}
\newblock \bibinfo{journal}{\bibinfo{title}{An internal model for sensorimotor
  integration}}.
\newblock {\emph{\JournalTitle{Science}}} \textbf{\bibinfo{volume}{269}},
  \bibinfo{pages}{1880--1882}, \doiprefix\url{10.1126/science.7569931}
  (\bibinfo{year}{1995}).

\bibitem{Brown2013}
\bibinfo{author}{Brown, H.}, \bibinfo{author}{Adams, R.~A.},
  \bibinfo{author}{Parees, I.}, \bibinfo{author}{Edwards, M.} \&
  \bibinfo{author}{Friston, K.}
\newblock \bibinfo{journal}{\bibinfo{title}{Active inference, sensory
  attenuation and illusions}}.
\newblock {\emph{\JournalTitle{Cognitive Processing}}}
  \textbf{\bibinfo{volume}{14}}, \bibinfo{pages}{411--427},
  \doiprefix\url{10.1007/s10339-013-0571-3} (\bibinfo{year}{2013}).

\bibitem{Friston2010}
\bibinfo{author}{Friston, K.}
\newblock \bibinfo{journal}{\bibinfo{title}{The free-energy principle: a
  unified brain theory?}}
\newblock {\emph{\JournalTitle{Nature Reviews Neuroscience}}}
  \textbf{\bibinfo{volume}{11}}, \bibinfo{pages}{127--138},
  \doiprefix\url{10.1038/nrn2787} (\bibinfo{year}{2010}).

\bibitem{Adams2013}
\bibinfo{author}{Adams, R.~A.}, \bibinfo{author}{Shipp, S.} \&
  \bibinfo{author}{Friston, K.~J.}
\newblock \bibinfo{journal}{\bibinfo{title}{Predictions not commands: active
  inference in the motor system}}.
\newblock {\emph{\JournalTitle{Brain Structure and Function}}}
  \textbf{\bibinfo{volume}{218}}, \bibinfo{pages}{611--643},
  \doiprefix\url{10.1007/s00429-012-0475-5} (\bibinfo{year}{2013}).

\bibitem{Ahmadi2019}
\bibinfo{author}{Ahmadi, A.} \& \bibinfo{author}{Tani, J.}
\newblock \bibinfo{journal}{\bibinfo{title}{A novel predictive-coding-inspired
  variational rnn model for online prediction and recognition}}.
\newblock {\emph{\JournalTitle{Neural Computation}}}
  \textbf{\bibinfo{volume}{31}}, \bibinfo{pages}{2025--2074},
  \doiprefix\url{10.1162/neco_a_01228} (\bibinfo{year}{2019}).

\bibitem{WOhata2020}
\bibinfo{author}{Ohata, W.} \& \bibinfo{author}{Tani, J.}
\newblock \bibinfo{journal}{\bibinfo{title}{Investigation of the sense of
  agency in social cognition, based on frameworks of predictive coding and
  active inference: A simulation study on multimodal imitative interaction}}.
\newblock {\emph{\JournalTitle{Frontiers in Neurorobotics}}}
  \textbf{\bibinfo{volume}{14:61}}, \doiprefix\url{10.3389/fnbot.2020.00061}
  (\bibinfo{year}{2020}).

\bibitem{Inoue2020}
\bibinfo{author}{Inoue, K.}, \bibinfo{author}{Nakajima, K.} \&
  \bibinfo{author}{Kuniyoshi, Y.}
\newblock \bibinfo{journal}{\bibinfo{title}{Designing spontaneous behavioral
  switching via chaotic itinerancy}}.
\newblock {\emph{\JournalTitle{Science Advances}}}
  \textbf{\bibinfo{volume}{6}}, \bibinfo{pages}{eabb3989},
  \doiprefix\url{10.1126/sciadv.abb3989} (\bibinfo{year}{2020}).

\bibitem{Haggard2004}
\bibinfo{author}{Haggard, P.} \& \bibinfo{author}{Whitford, B.}
\newblock \bibinfo{journal}{\bibinfo{title}{Supplementary motor area provides
  an efferent signal for sensory suppression}}.
\newblock {\emph{\JournalTitle{Cognitive Brain Research}}}
  \textbf{\bibinfo{volume}{19}}, \bibinfo{pages}{52--58},
  \doiprefix\url{10.1016/j.cogbrainres.2003.10.018} (\bibinfo{year}{2004}).

\bibitem{Wolpe2016}
\bibinfo{author}{Wolpe, N.} \emph{et~al.}
\newblock \bibinfo{journal}{\bibinfo{title}{Ageing increases reliance on
  sensorimotor prediction through structural and functional differences in
  frontostriatal circuits}}.
\newblock {\emph{\JournalTitle{Nature Communications}}}
  \textbf{\bibinfo{volume}{7}}, \bibinfo{pages}{13034},
  \doiprefix\url{10.1038/ncomms13034} (\bibinfo{year}{2016}).

\bibitem{Boehme2019}
\bibinfo{author}{Boehme, R.}, \bibinfo{author}{Hauser, S.},
  \bibinfo{author}{Gerling, G.~J.}, \bibinfo{author}{Heilig, M.} \&
  \bibinfo{author}{Olausson, H.}
\newblock \bibinfo{journal}{\bibinfo{title}{Distinction of self-produced touch
  and social touch at cortical and spinal cord levels}}.
\newblock {\emph{\JournalTitle{Proceedings of the National Academy of
  Sciences}}} \textbf{\bibinfo{volume}{116}}, \bibinfo{pages}{2290},
  \doiprefix\url{10.1073/pnas.1816278116} (\bibinfo{year}{2019}).

\bibitem{Eagleman2004}
\bibinfo{author}{Eagleman, D.~M.}
\newblock \bibinfo{journal}{\bibinfo{title}{The where and when of intention}}.
\newblock {\emph{\JournalTitle{Science}}} \textbf{\bibinfo{volume}{303}},
  \bibinfo{pages}{1144--1146}, \doiprefix\url{10.1126/science.1095331}
  (\bibinfo{year}{2004}).

\bibitem{Leek2009}
\bibinfo{author}{Leek, E.~C.} \& \bibinfo{author}{Johnston, S.~J.}
\newblock \bibinfo{journal}{\bibinfo{title}{Functional specialization in the
  supplementary motor complex}}.
\newblock {\emph{\JournalTitle{Nature Reviews Neuroscience}}}
  \textbf{\bibinfo{volume}{10}}, \bibinfo{pages}{78--78},
  \doiprefix\url{10.1038/nrn2478-c1} (\bibinfo{year}{2009}).

\bibitem{Yu2005}
\bibinfo{author}{Yu, A.~J.} \& \bibinfo{author}{Dayan, P.}
\newblock \bibinfo{journal}{\bibinfo{title}{Uncertainty, neuromodulation, and
  attention}}.
\newblock {\emph{\JournalTitle{Neuron}}} \textbf{\bibinfo{volume}{46}},
  \bibinfo{pages}{681^^e2^^80^^93692},
  \doiprefix\url{10.1016/j.neuron.2005.04.026} (\bibinfo{year}{2005}).

\bibitem{Corlett2010}
\bibinfo{author}{Corlett, P.}, \bibinfo{author}{Taylor, J.},
  \bibinfo{author}{Wang, X.-J.}, \bibinfo{author}{Fletcher, P.} \&
  \bibinfo{author}{Krystal, J.}
\newblock \bibinfo{journal}{\bibinfo{title}{Toward a neurobiology of
  delusions}}.
\newblock {\emph{\JournalTitle{Progress in Neurobiology}}}
  \textbf{\bibinfo{volume}{92}}, \bibinfo{pages}{345--369},
  \doiprefix\url{10.1016/j.pneurobio.2010.06.007} (\bibinfo{year}{2010}).

\bibitem{Pynn2013}
\bibinfo{author}{Pynn, L.~K.} \& \bibinfo{author}{DeSouza, J.~F.}
\newblock \bibinfo{journal}{\bibinfo{title}{The function of efference copy
  signals: Implications for symptoms of schizophrenia}}.
\newblock {\emph{\JournalTitle{Vision Research}}}
  \textbf{\bibinfo{volume}{76}}, \bibinfo{pages}{124--133},
  \doiprefix\url{10.1016/j.visres.2012.10.019} (\bibinfo{year}{2013}).

\bibitem{Kilteni2022}
\bibinfo{author}{Kilteni, K.} \& \bibinfo{author}{Ehrsson, H.~H.}
\newblock \bibinfo{journal}{\bibinfo{title}{Predictive attenuation of touch and
  tactile gating are distinct perceptual phenomena}}.
\newblock {\emph{\JournalTitle{iScience}}} \textbf{\bibinfo{volume}{25}},
  \bibinfo{pages}{104077},
  \doiprefix\url{https://doi.org/10.1016/j.isci.2022.104077}
  (\bibinfo{year}{2022}).

\bibitem{Kilteni2019}
\bibinfo{author}{Kilteni, K.}, \bibinfo{author}{Houborg, C.} \&
  \bibinfo{author}{Ehrsson, H.~H.}
\newblock \bibinfo{journal}{\bibinfo{title}{Rapid learning and unlearning of
  predicted sensory delays in^^c2^^a0self-generated touch}}.
\newblock {\emph{\JournalTitle{eLife}}} \textbf{\bibinfo{volume}{8}},
  \bibinfo{pages}{e42888}, \doiprefix\url{10.7554/eLife.42888}
  (\bibinfo{year}{2019}).

\bibitem{Palmer2016}
\bibinfo{author}{Palmer, C.~E.}, \bibinfo{author}{Davare, M.} \&
  \bibinfo{author}{Kilner, J.~M.}
\newblock \bibinfo{journal}{\bibinfo{title}{Physiological and perceptual
  sensory attenuation have different underlying neurophysiological
  correlates}}.
\newblock {\emph{\JournalTitle{Journal of Neuroscience}}}
  \textbf{\bibinfo{volume}{36}}, \bibinfo{pages}{10803--10812},
  \doiprefix\url{10.1523/JNEUROSCI.1694-16.2016} (\bibinfo{year}{2016}).
\newblock \eprint{https://www.jneurosci.org/content/36/42/10803.full.pdf}.

\bibitem{Thomas2022}
\bibinfo{author}{Thomas, E.~R.}, \bibinfo{author}{Yon, D.},
  \bibinfo{author}{de~Lange, F.~P.} \& \bibinfo{author}{Press, C.}
\newblock \bibinfo{journal}{\bibinfo{title}{Action enhances predicted touch}}.
\newblock {\emph{\JournalTitle{Psychological Science}}}
  \textbf{\bibinfo{volume}{33}}, \bibinfo{pages}{48--59},
  \doiprefix\url{10.1177/09567976211017505} (\bibinfo{year}{2022}).
\newblock \bibinfo{note}{PMID: 34878943},
  \eprint{https://doi.org/10.1177/09567976211017505}.

\bibitem{Powers2017}
\bibinfo{author}{Powers, A.~R.}, \bibinfo{author}{Mathys, C.} \&
  \bibinfo{author}{Corlett, P.~R.}
\newblock \bibinfo{journal}{\bibinfo{title}{Pavlovian
  conditioning^^e2^^80^^93induced hallucinations result from overweighting of
  perceptual priors}}.
\newblock {\emph{\JournalTitle{Science}}} \textbf{\bibinfo{volume}{357}},
  \bibinfo{pages}{596--600}, \doiprefix\url{10.1126/science.aan3458}
  (\bibinfo{year}{2017}).

\bibitem{Corlett2019}
\bibinfo{author}{Corlett, P.~R.} \emph{et~al.}
\newblock \bibinfo{journal}{\bibinfo{title}{Hallucinations and strong priors}}.
\newblock {\emph{\JournalTitle{Trends in Cognitive Sciences}}}
  \textbf{\bibinfo{volume}{23}}, \bibinfo{pages}{114--127},
  \doiprefix\url{10.1016/j.tics.2018.12.001} (\bibinfo{year}{2019}).

\bibitem{Blakemore2006}
\bibinfo{author}{Blakemore, S.-J.} \emph{et~al.}
\newblock \bibinfo{journal}{\bibinfo{title}{Tactile sensitivity in asperger
  syndrome}}.
\newblock {\emph{\JournalTitle{Brain and Cognition}}}
  \textbf{\bibinfo{volume}{61}}, \bibinfo{pages}{5--13},
  \doiprefix\url{10.1016/j.bandc.2005.12.013} (\bibinfo{year}{2006}).

\bibitem{Lawson2014}
\bibinfo{author}{Lawson, R.~P.}, \bibinfo{author}{Rees, G.} \&
  \bibinfo{author}{Friston, K.~J.}
\newblock \bibinfo{journal}{\bibinfo{title}{{An aberrant precision account of
  autism}}}.
\newblock {\emph{\JournalTitle{Front. Hum. Neurosci.}}}
  \textbf{\bibinfo{volume}{8}}, \bibinfo{pages}{302},
  \doiprefix\url{10.3389/fnhum.2014.00302} (\bibinfo{year}{2014}).

\bibitem{Haker2016}
\bibinfo{author}{Haker, H.}, \bibinfo{author}{Schneebeli, M.} \&
  \bibinfo{author}{Stephan, K.~E.}
\newblock \bibinfo{journal}{\bibinfo{title}{Can bayesian theories of autism
  spectrum disorder help improve clinical practice?}}
\newblock {\emph{\JournalTitle{Frontiers in Psychiatry}}}
  \textbf{\bibinfo{volume}{7}} (\bibinfo{year}{2016}).

\bibitem{Palmer2017}
\bibinfo{author}{Palmer, C.~J.}, \bibinfo{author}{Lawson, R.~P.} \&
  \bibinfo{author}{Hohwy, J.}
\newblock \bibinfo{journal}{\bibinfo{title}{Bayesian approaches to autism:
  Towards volatility, action, and behavior.}}
\newblock {\emph{\JournalTitle{Psychological Bulletin}}}
  \textbf{\bibinfo{volume}{143}}, \bibinfo{pages}{521--542},
  \doiprefix\url{10.1037/bul0000097} (\bibinfo{year}{2017}).

\bibitem{Finnemann2021}
\bibinfo{author}{Finnemann, J.~J.}, \bibinfo{author}{Plaisted-Grant, K.},
  \bibinfo{author}{Moore, J.}, \bibinfo{author}{Teufel, C.} \&
  \bibinfo{author}{Fletcher, P.~C.}
\newblock \bibinfo{journal}{\bibinfo{title}{Low-level, prediction-based sensory
  and motor processes are unimpaired in autism}}.
\newblock {\emph{\JournalTitle{Neuropsychologia}}}
  \textbf{\bibinfo{volume}{156}}, \bibinfo{pages}{107835},
  \doiprefix\url{10.1016/j.neuropsychologia.2021.107835}
  (\bibinfo{year}{2021}).

\bibitem{Yamashita2008}
\bibinfo{author}{Yamashita, Y.} \& \bibinfo{author}{Tani, J.}
\newblock \bibinfo{journal}{\bibinfo{title}{{Emergence of functional hierarchy
  in a multiple timescale neural network model: A humanoid robot experiment}}}.
\newblock {\emph{\JournalTitle{PLoS Comput. Biol.}}}
  \textbf{\bibinfo{volume}{4}}, \bibinfo{pages}{e1000220}
  (\bibinfo{year}{2008}).

\bibitem{Yamashita2012}
\bibinfo{author}{Yamashita, Y.} \& \bibinfo{author}{Tani, J.}
\newblock \bibinfo{journal}{\bibinfo{title}{{Spontaneous prediction error
  generation in schizophrenia}}}.
\newblock {\emph{\JournalTitle{PLoS ONE}}} \textbf{\bibinfo{volume}{7}},
  \bibinfo{pages}{e37843--e37843}, \doiprefix\url{10.1371/journal.pone.0037843}
  (\bibinfo{year}{2012}).

\bibitem{Idei2018}
\bibinfo{author}{Idei, H.} \emph{et~al.}
\newblock \bibinfo{journal}{\bibinfo{title}{A neurorobotics simulation of
  autistic behavior induced by unusual sensory precision}}.
\newblock {\emph{\JournalTitle{Computational Psychiatry}}}
  \textbf{\bibinfo{volume}{2}}, \bibinfo{pages}{164--182},
  \doiprefix\url{10.1162/CPSY_a_00019} (\bibinfo{year}{2018}).

\bibitem{Idei2020}
\bibinfo{author}{Idei, H.}, \bibinfo{author}{Murata, S.},
  \bibinfo{author}{Yamashita, Y.} \& \bibinfo{author}{Ogata, T.}
\newblock \bibinfo{journal}{\bibinfo{title}{Homogeneous intrinsic neuronal
  excitability induces overfitting to sensory noise: A robot model of
  neurodevelopmental disorder}}.
\newblock {\emph{\JournalTitle{Frontiers in Psychiatry}}}
  \textbf{\bibinfo{volume}{11:762}}, \doiprefix\url{10.3389/fpsyt.2020.00762}
  (\bibinfo{year}{2020}).

\bibitem{Idei2021}
\bibinfo{author}{Idei, H.}, \bibinfo{author}{Murata, S.},
  \bibinfo{author}{Yamashita, Y.} \& \bibinfo{author}{Ogata, T.}
\newblock \bibinfo{journal}{\bibinfo{title}{Paradoxical sensory reactivity
  induced by functional disconnection in a robot model of neurodevelopmental
  disorder}}.
\newblock {\emph{\JournalTitle{Neural Networks}}}
  \textbf{\bibinfo{volume}{138}}, \bibinfo{pages}{150--163},
  \doiprefix\url{10.1016/j.neunet.2021.01.033} (\bibinfo{year}{2021}).

\bibitem{Finkelstein2021}
\bibinfo{author}{Finkelstein, A.} \emph{et~al.}
\newblock \bibinfo{journal}{\bibinfo{title}{Attractor dynamics gate cortical
  information flow during decision-making}}.
\newblock {\emph{\JournalTitle{Nature Neuroscience}}}
  \textbf{\bibinfo{volume}{24}}, \bibinfo{pages}{843--850},
  \doiprefix\url{10.1038/s41593-021-00840-6} (\bibinfo{year}{2021}).

\bibitem{Newell2001}
\bibinfo{author}{Newell, K.}, \bibinfo{author}{Liu, Y.} \&
  \bibinfo{author}{Mayer-Kress, G.}
\newblock \bibinfo{journal}{\bibinfo{title}{Time scales in motor learning and
  development.}}
\newblock {\emph{\JournalTitle{Psychological review}}}
  \textbf{\bibinfo{volume}{108 1}}, \bibinfo{pages}{57--82}
  (\bibinfo{year}{2001}).

\bibitem{Kingma2017}
\bibinfo{author}{Kingma, D.~P.} \& \bibinfo{author}{Ba, J.}
\newblock \bibinfo{title}{Adam: A method for stochastic optimization}
  (\bibinfo{year}{2017}).
\newblock \eprint{1412.6980}.

\bibitem{Azouz2000}
\bibinfo{author}{Azouz, R.} \& \bibinfo{author}{Gray, C.~M.}
\newblock \bibinfo{journal}{\bibinfo{title}{Dynamic spike threshold reveals a
  mechanism for synaptic coincidence detection in cortical neurons {\em{in
  vivo}}}}.
\newblock {\emph{\JournalTitle{Proceedings of the National Academy of
  Sciences}}} \textbf{\bibinfo{volume}{97}}, \bibinfo{pages}{8110--8115},
  \doiprefix\url{10.1073/pnas.130200797} (\bibinfo{year}{2000}).

\bibitem{Mejias2012}
\bibinfo{author}{Mejias, J.~F.} \& \bibinfo{author}{Longtin, A.}
\newblock \bibinfo{journal}{\bibinfo{title}{Optimal heterogeneity for coding in
  spiking neural networks}}.
\newblock {\emph{\JournalTitle{Physical Review Letters}}}
  \textbf{\bibinfo{volume}{108}}, \bibinfo{pages}{228102--1--5},
  \doiprefix\url{10.1103/PhysRevLett.108.228102} (\bibinfo{year}{2012}).

\end{thebibliography}




\section*{Acknowledgements}
This work was partially supported by an unrestricted gift from Google and supported by a JSPS Grant-in-Aid (Nos. JP19J20281, JP22J01708), JST Moonshot R\&D (No. JPMJMS2031), and JST CREST grants (Nos. JPMJCR16E2, JPMJCR21P4).%

\section*{Author contributions statement}
Conceptualization: HI, JT \\
\quad Methodology: HI, YY, JT \\
\quad Investigation: HI, WO \\
\quad Visualization: HI, YY, JT \\
\quad Funding acquisition: HI, YY, TO, JT \\
\quad Project administration: JT \\
\quad Supervision: JT \\
\quad Writing -- original draft: HI \\
\quad Writing -- review \& editing: HI, WO, YY, TO, JT \\

\section*{Additional information}
\subsection*{Data Availability:}
All data is available in the manuscript and the supplementary information. %
\subsection*{Code Availability:}
Computer code for the neural network model was written using Pytorch (a library for deep learning) and is available online (\url{https://github.com/h-idei/pvrnn_sa.git}). %
\subsection*{Competing interests:} 
Authors declare that they have no competing interests. %

\clearpage
\renewcommand{\figurename}{Figure}

\begin{figure}[ht]
  \centering
  \includegraphics[width=\textwidth]{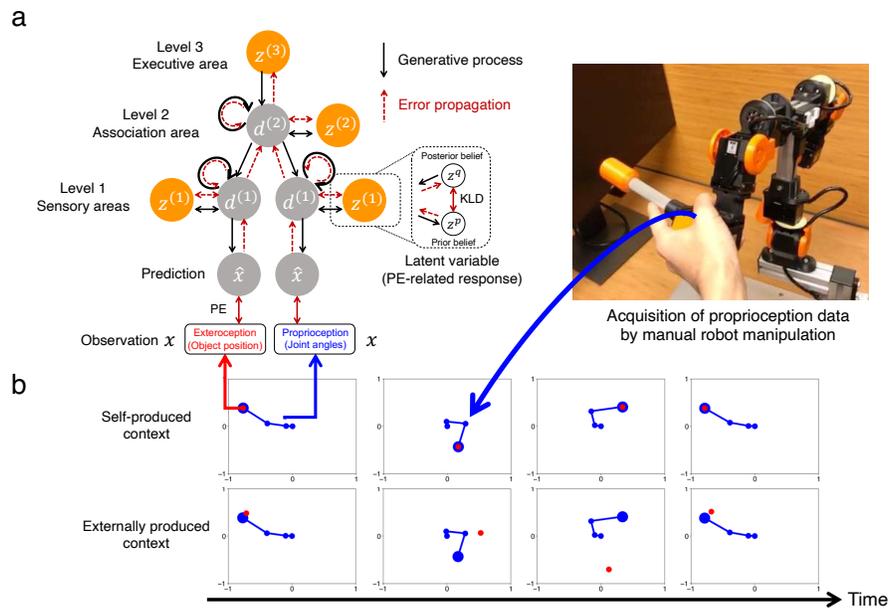} 
  \caption{Training setting. {\bf (a)} Hierarchical recurrent network updating of synaptic weights and posterior beliefs in all areas via minimization of free-energy determined by the prediction error (PE) and Kullback-Leibler divergence (KLD) between the posterior and the prior. {\bf (b)} Spontaneous sensorimotor patterns learned by the recurrent network.} 
  \label{Fig1}
\end{figure}

\begin{figure}[ht]
  \centering
  \includegraphics[width=\textwidth]{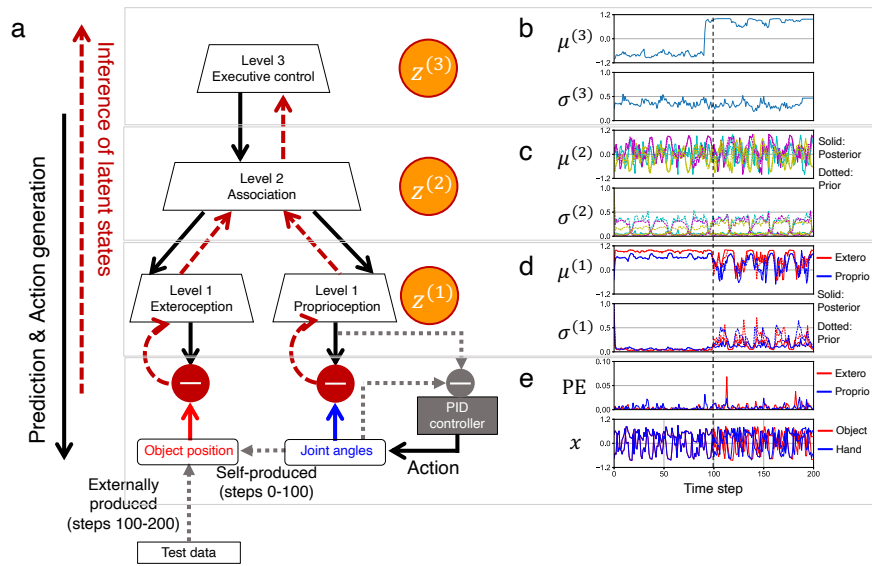}
  \caption{Test setting and an example of the result. {\bf (a)} Online inference of latent states and action generation by a PID controller, with synaptic weights fixed. {\bf (b)} Executive-level posterior distribution. {\bf (c)} Association-level posterior and prior distributions. {\bf (d)} Sensory-level posterior and prior distributions in exteroceptive and proprioceptive areas. {\bf (e)} Prediction error (PE) and real sensations for exteroception and proprioception. For clarity, proprioception is represented as the 2-dimensional hand position, although the actual proprioception comprises 3-dimensional joint angles.}
  \label{Fig2}
\end{figure}

\begin{figure}[ht]
  \centering
  \includegraphics[width=\textwidth]{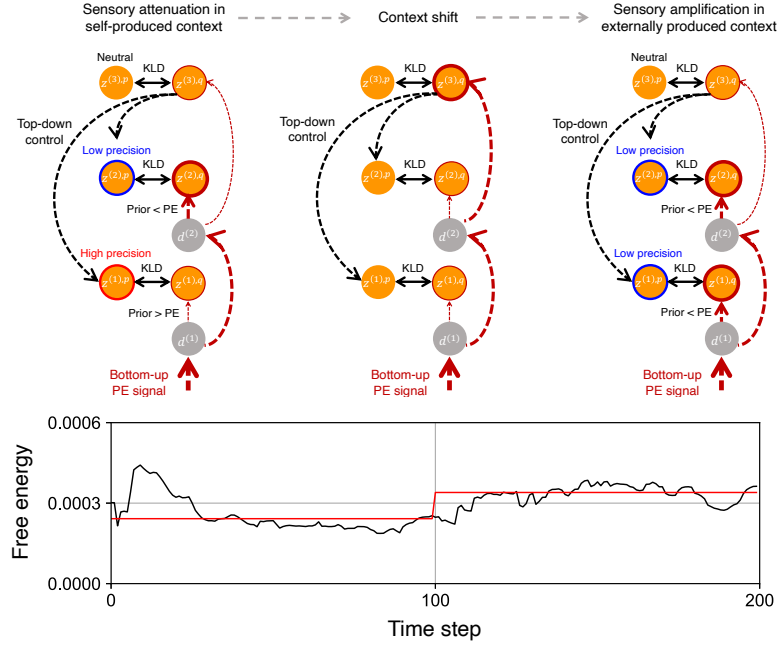}
  \caption{The self-organized mechanism for the transition between sensory attenuation in self-produced contexts and sensory amplification in externally produced contexts. For each sensorimotor context, a particular free-energy minimum in the network was developed through learning, which was characterized by the corresponding top-down precision control and bottom-up prediction-error (PE) flow. Consequently, in self-produced contexts, the recurrent network mainly adjusted only the posterior in the association area $\bm{z^{(2),q}}$ to minimize prediction errors by increasing precision of priors in sensory areas $\bm{z^{(1),p}}$ and decreasing precision of prior in the association area $\bm{z^{(2),p}}$. On the other hand, in externally produced contexts, it adjusted posteriors in both sensory and association areas (\bm{$z^{(1),q}}$ and $\bm{z^{(2),q}}$) by decreasing precision of priors in both sensory and association areas. The shift of the sensorimotor context triggered modulation of the posterior in the executive area $\bm{z^{(3),q}}$ as well as the transition from one free-energy state to another in the network (bottom time series). This induced a qualitative change in the network behavior and a shift between attenuating and amplifying prediction-error-induced responses in sensory areas. The black line in the time series of the free-energy describes the 20-step moving average (values within the first 19 steps are calculated as averages from the initial to the current step). The red line is the average over the first 100 steps, i.e., self-produced context, or the last 100 steps, i.e., externally produced context. Values are averages over 80 test trials performed by 10 different networks.}
  \label{Fig3}
\end{figure}

\begin{figure}[ht]
  \centering
  \includegraphics[width=\textwidth]{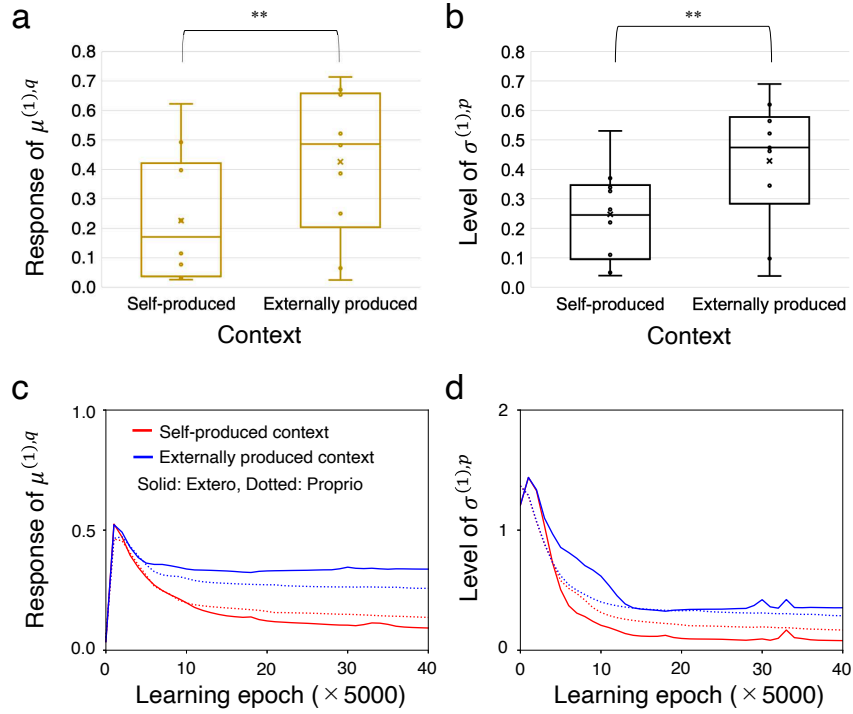}
  \caption{Sensory-level neural response in the test phase and its development in the learning phase. {\bf (a)} Box-and-whiskers plots showing the difference in sensory-level posterior response between self-produced and externally produced contexts. {\bf (b)} Box-and-whiskers plots showing the difference in the sensory-level prior sigma between the two contexts. In (a) and (b), each plot is an average of 100 time steps, 2 sensory areas, and 8 test trials by each of 10 trained networks (center line, median; cross, mean; box limits, upper and lower quartiles; whiskers, 1.5x interquartile range). **$p<0.01$. {\bf (c)} Development of the posterior response for the two contexts in the exteroceptive and proprioceptive areas. {\bf (d)} Development of sensory-level prior sigma. In (c) and (d), the value is averaged over 200 time steps, 24 training datasets, and 10 different networks in each learning epoch.} 
  \label{Fig4}
\end{figure}

\clearpage

\section*{Supplementary Figures:}
\setcounter{figure}{0}
\renewcommand{\figurename}{Supplementary Figure S}

\begin{figure}[ht]
  \centering
  \includegraphics[width=\textwidth]{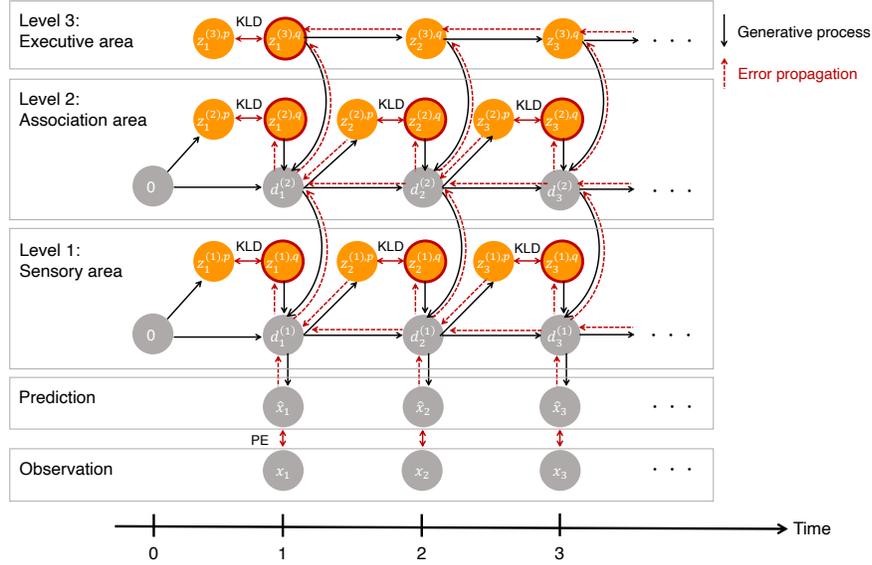}
  \caption{Temporal processing of the recurrent neural network. In the learning phase, the recurrent network updates the posteriors of latent variables in all areas ($\bm{z_{1:T}^{(1),q}}$, $\bm{z_{1:T}^{(2),q}}$, and $\bm{z_{1}^{(3),q}}$) and time-constant synaptic weights via minimization of free-energy over the time length (T) of training data. Initial deterministic states $\bm{d_{0}}$ in all areas are set to $0$. The initial prior distribution $\bm{z_{1}^{p}}$ in the executive area is set to a unit Gaussian distribution $\mathcal{N}(0,1)$. For simplicity, the distributed structure of sensory areas is omitted. PE: Prediction error. KLD: Kullback-Leibler divergence between the posterior and the prior of latent variable.} 
  \label{FigS1}
\end{figure}

\clearpage

\begin{figure}[ht]
  \centering
  \includegraphics[width=\textwidth]{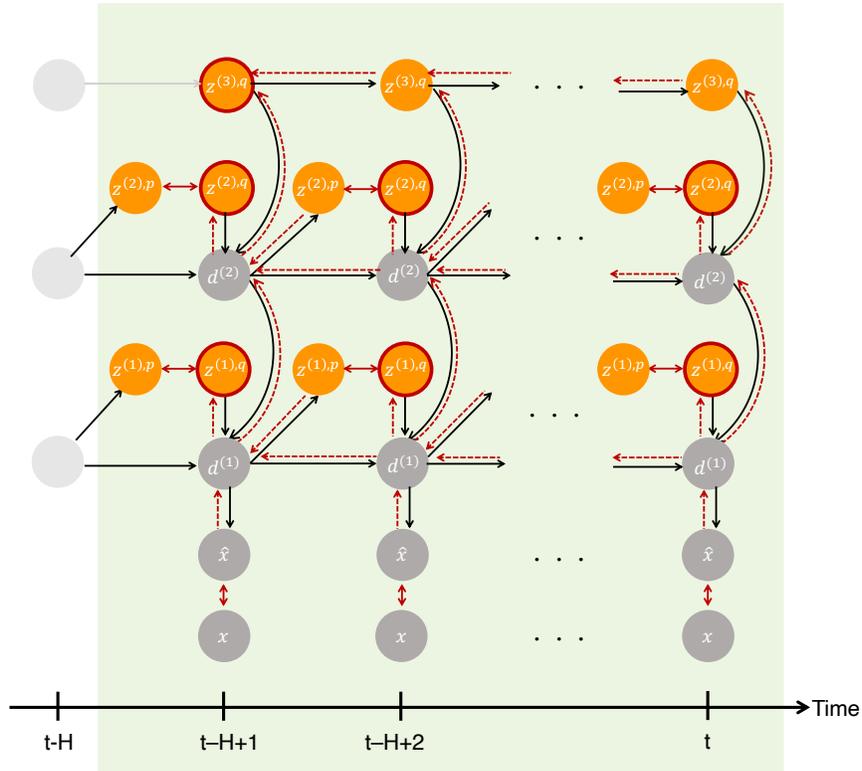}
  \caption{Online inference in the test phase. The recurrent network updates the posterior ($\bm{z_{t-H+1:t}^{(1),q}}$, $\bm{z_{t-H+1:t}^{(2),q}}$, and $\bm{z_{t-H+1}^{(3),q}}$) via minimization of the free-energy over a certain time window (H). Synaptic weights are fixed. For simplicity, the distributed structure of sensory areas is omitted.} 
  \label{FigS2}
\end{figure}

\begin{figure}[ht]
  \centering
  \includegraphics[width=\textwidth]{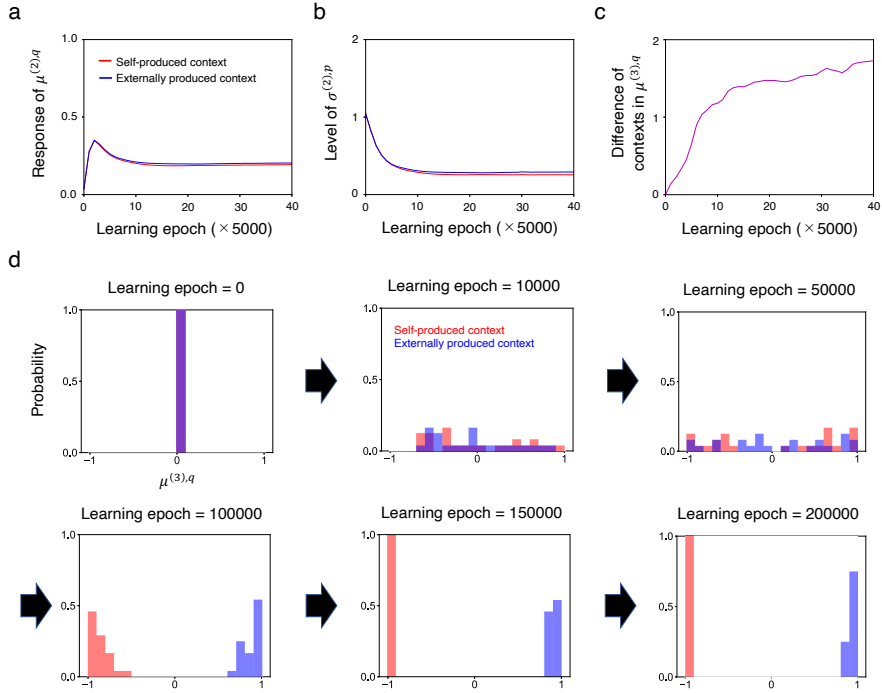}
  \caption{Development of higher-level latent variables. {\bf (a-b)} Development of a posterior response and a prior sigma in the association area. At each learning epoch, the value is averaged over 200 time steps, 24 training datasets, and 10 networks. {\bf (c)} Development of the difference between self-produced contexts and externally produced contexts in the executive area. In the learning phase, the recurrent network developed a posterior state in the executive area for each of 48 training datasets (24 for self-produced contexts and 24 for externally produced contexts). The figure shows the distance between median values of posterior states for self-produced and externally produced contexts. At each learning epoch, the value is averaged over 10 networks. {\bf (d)} An example of development of the executive-level posterior. Each figure shows the distribution of 24 executive-level posterior states corresponding to self-produced or externally produced contexts. The figure shows that recognition of different sensorimotor experiences was gradually developed through a learning process. Note that we did not provide any explicit label indicating the difference between the two contexts.} 
  \label{FigS3}
\end{figure}

\begin{figure}[ht]
  \centering
  \includegraphics[width=\textwidth]{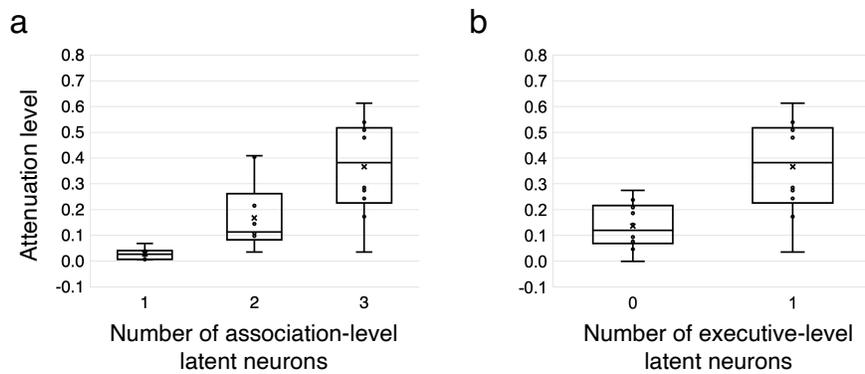}
  \caption{Neural-response attenuation varying with the number of latent neurons. The attenuation level was calculated by subtracting the sensory-level posterior response in a self-produced context from that in an externally produced context while reproducing training data. Values were for 10 trained networks with different initial synaptic weights. {\bf (a)} A larger number of association-level latent neurons led to a larger attenuation level, suggesting that representing sensorimotor correlation required enough association-level latent neurons. {\bf (b)} When there were no executive-level latent neurons, the attenuation level was greatly reduced. This suggests the importance of executive-level control for sensory attenuation. In (a) and (b), the setting of 3 association-level and 1 executive-level latent neurons was the same baseline model.} 
  \label{FigS4}
\end{figure}

\begin{figure}[ht]
  \centering
  \includegraphics[width=\textwidth]{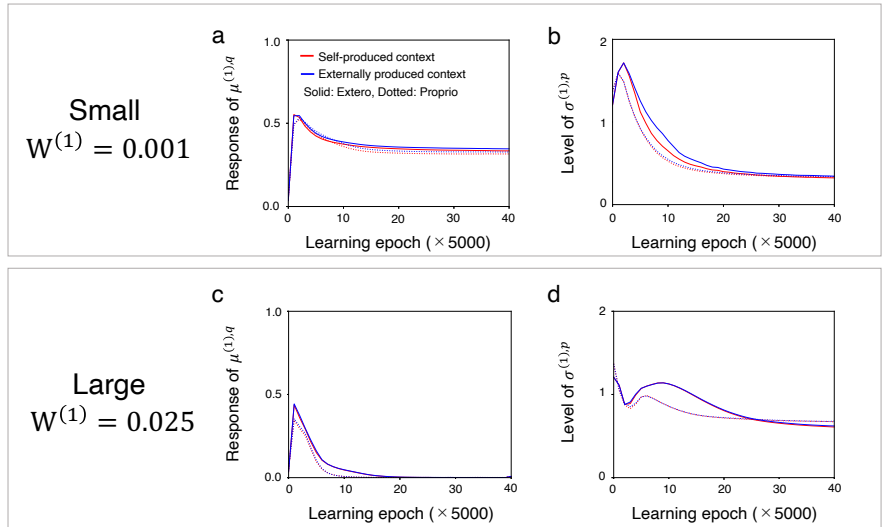}
  \caption{Effects of shifts in the meta-prior in sensory areas.  {\bf (a-b)} Development of a sensory-level posterior response and prior sigma through learning in a small sensory-level meta-prior setting ($W^{(1)}=0.001, W^{(2)}=W^{(3)}=0.005$), described as in Fig. 4c-d. A small sensory-level meta-prior led to reduced attenuation of the sensory-level posterior response, as well as the sensory-level prior sigma, in the self-produced context. {\bf (c-d)} Development of a sensory-level posterior response and a prior sigma in a large sensory-level meta-prior setting ($W^{(1)}=0.025, W^{(2)}=W^{(3)}=0.005$). A large sensory-level meta-prior led to highly reduced sensory-level posterior responses in both self-produced and externally produced contexts.} 
  \label{FigS5}
\end{figure}

\begin{figure}[ht]
  \centering
  \includegraphics[width=\textwidth]{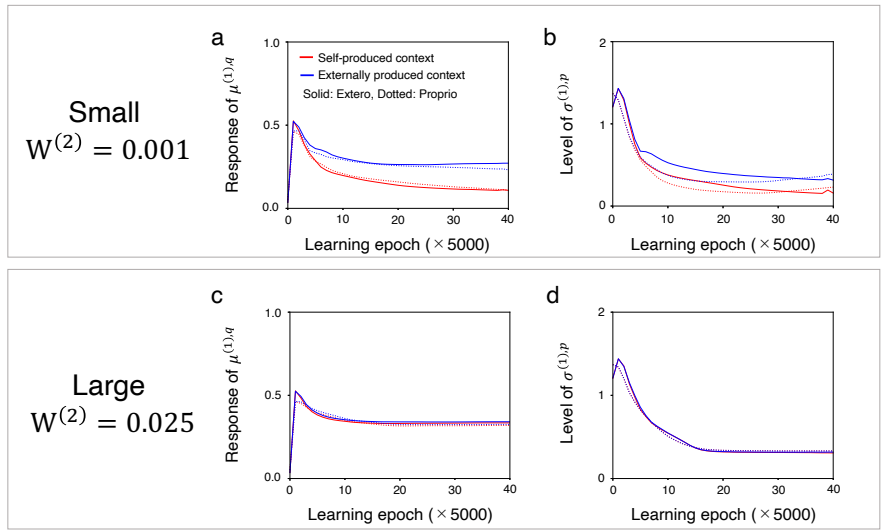}
  \caption{Effects of shifts in the meta-prior in the association area. {\bf (a-b)} Development of a sensory-level posterior response and prior sigma through learning in a small association-level meta-prior setting ($W^{(2)}=0.001, W^{(1)}=W^{(3)}=0.005$), described as in Fig. 4c-d. A small association-level meta-prior did not have a large impact on development of sensory attenuation. {\bf (c-d)} Development of a sensory-level posterior response and a prior sigma in a large association-level meta-prior setting ($W^{(2)}=0.025, W^{(1)}=W^{(3)}=0.005$). A large association-level meta-prior led to reduced attenuation of the sensory-level posterior response, as well as the sensory-level prior sigma, in the self-produced context.} 
  \label{FigS6}
\end{figure}

\begin{figure}[ht]
  \centering
  \includegraphics[width=\textwidth]{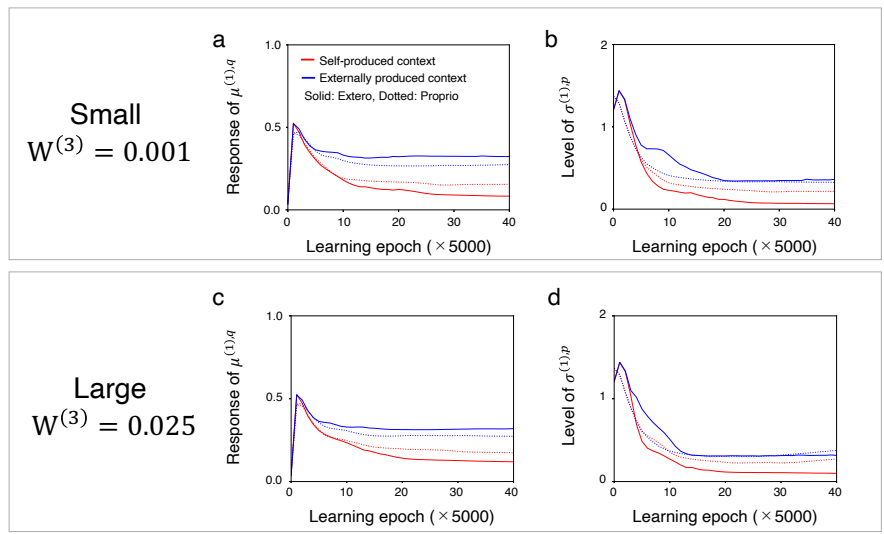}
  \caption{Effects of shifts in the meta-prior in the executive area. {\bf (a-b)}Development of a sensory-level posterior response and prior sigma through learning in a small executive-level meta-prior setting ($W^{(3)}=0.001, W^{(1)}=W^{(2)}=0.005$), described as in Fig. 4c-d. {\bf (c-d)} Development of a sensory-level posterior response and a prior sigma in a large executive-level meta-prior setting ($W^{(3)}=0.025, W^{(1)}=W^{(2)}=0.005$). Neither small nor large executive-level meta-priors had a large impact on development of the sensory-level posterior response or the prior sigma.} 
  \label{FigS7}
\end{figure}

\begin{figure}[ht]
  \centering
  \includegraphics[width=\textwidth]{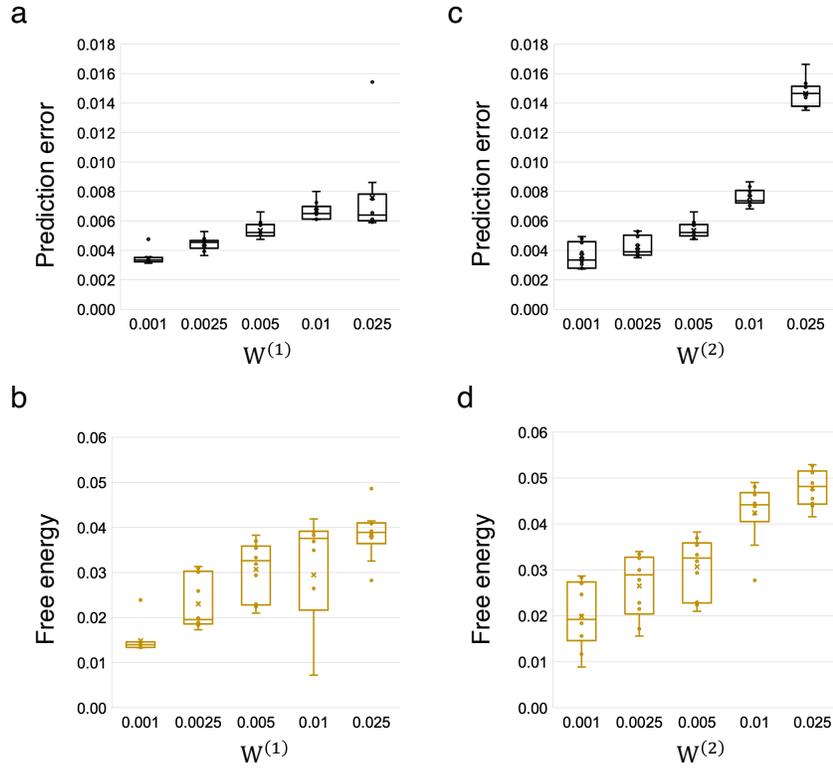}
  \caption{Training results. {\bf (a-b)} Prediction error and free-energy for training data under each meta-prior setting in sensory areas ($W^{(2)}=W^{(3)}=0.005$). A small sensory-level meta-prior led to decreases in the prediction error and free-energy, although development of sensory attenuation was disrupted. {\bf (c-d)} The prediction error and free-energy for training data under each meta-prior setting in the association area ($W^{(1)}=W^{(3)}=0.005$). In (a)-(d), the setting of $W^{(1)}=W^{(2)}=W^{(3)}=0.005$ was the same baseline model.} 
  \label{FigS8}
\end{figure}

\begin{figure}[ht]
  \centering
  \includegraphics[width=\textwidth]{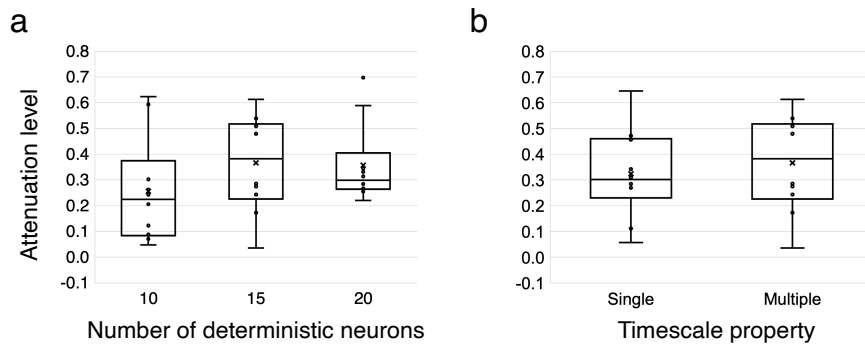}
  \caption{Neural-response attenuation varied with settings of deterministic neurons. The attenuation level was calculated by subtracting the sensory-level posterior response in the self-produced context from that in the externally produced context while reproducing training data. Values were for 10 trained networks with different initial synaptic weights. {\bf (a)} A recurrent network with 15 deterministic neurons showed the largest average (and median) attenuation level. {\bf (b)} Multiple timescale settings ($\tau=2$ for 8 neurons and $\tau=4$ for 7 neurons) increased the attenuation level compared to the single timescale setting ($\tau=2$ for all deterministic neurons). In (a) and (b), the setting of 15 deterministic neurons and multiple timescale property was the same as the baseline model.} 
  \label{FigS9}
\end{figure}

\clearpage
\section*{Caption for Supplementary Video:}
\noindent
Supplementary Video S1. Robot behavior and neural responses in the baseline model. Related to Fig. 2b-e. Red and blue crosses in the left image indicate predicted positions of the external object and robot hand, respectively. \\\\


\end{document}